\documentclass[journal]{IEEEtran}

\usepackage{tabularx} 
\usepackage{graphicx}
\usepackage{xcolor}
\definecolor{RED}{named}{red}
\usepackage{fancyhdr}
\PassOptionsToPackage{hyphens}{url}\usepackage{hyperref}

\begin{document}
\bstctlcite{IEEEexample:BSTcontrol}

\title{A Room With an Overview: Towards Meaningful Transparency for the Consumer Internet of Things}

\author{Chris~Norval
        and~Jatinder~Singh%
\thanks{The authors are part of the Compliant \& Accountable Systems Group at the University of Cambridge, William Gates Building, Cambridge, CB3 0FD, United Kingdom e-mail: (cjn41@cam.ac.uk; jatinder.singh@cl.cam.ac.uk).}}%

\maketitle

\pagestyle{fancy}
\fancyhead{}
\fancyhead[L]{
\begin{scriptsize}\color{red}To Appear: \hspace {.05cm} 
C. Norval and J. Singh, "A Room With an Overview: Towards Meaningful Transparency for the Consumer Internet of Things," in IEEE Internet of Things Journal. DOI: \href{https://doi.org/10.1109/JIOT.2023.3318369}{10.1109/JIOT.2023.3318369}.
\end{scriptsize}
}

\begin{abstract}
As our physical environments become ever-more connected, instrumented and automated, it can be increasingly difficult for users to understand what is happening within them and why. 
This warrants attention; with the pervasive and physical nature of the IoT comes risks of data misuse, privacy, surveillance, and even physical harm.
Such concerns come amid increasing calls for more transparency surrounding technologies (in general), as a means for supporting scrutiny and accountability.
This paper explores the practical dimensions to transparency mechanisms within the consumer IoT.
That is, we consider how smart homes might be made more \textit{meaningfully transparent}, so as to support users in gaining greater understanding, oversight, and control.
Through a series of three user-centric studies, we (i) survey prospective smart home users to gain a general understanding of what meaningful transparency within smart homes might entail; (ii) identify categories of user-derived requirements and design elements (design features for supporting smart home transparency) that have been created through two co-design workshops; and (iii) validate these  through an evaluation with an altogether new set of participants.
In all, these categories of requirements and interface design elements provide a foundation for understanding how meaningful transparency might be achieved within smart homes, and introduces several wider considerations for doing so.
\end{abstract}


\IEEEpeerreviewmaketitle

\section{Introduction} \label{sec:intro}

Our environments are becoming increasingly connected, instrumented, and automated, amid an ever-growing myriad of network-enabled consumer Internet of Things (IoT) devices claiming to offer convenience, comfort, safety, and control~\cite{chalhoub2021, davidoff2006, jakobi2017, zheng2018}. 
From lighting, heating, and appliances to home security and beyond, surveys indicate that the average US household has eleven connected devices, with 28\% of households having at least one home automation device~\cite{deloitte2021}.
In this way, the IoT is already impacting the lives of many.

Most consumer IoT devices typically have domestic and lifestyle aims, enabling interactions between the user and their surroundings.
For example, a smart thermostat might work to automatically manage the temperature of a home, turning the heating on when it is cold, or opening the windows if warm.
However, the inner workings (i.e. the operation) of these systems can quickly become complex and opaque; a given deployment might entail numerous interactions (and data flows) between a range of components, where, for example, even fairly simple automated events (such as smart windows opening in response to weather reports) can be driven by a supply chain of data sources and dependencies~\cite{singh2019}. 
In short, what data is collected, how it is used, and where it goes are often little-known by users~\cite{jakobi2018, tabassum2019}, and are typically obscured within and as part of the broader IoT ecosystem~\cite{norval2021}.
This hinders users in understanding and overseeing what is going on within their IoT deployment (which they may do so for reasons such as verifying that the deployment is operating as expected, in response to particular incidents or events, or for general curiosity).
As a result, it may not be clear why the system operates in the way that it does, potentially leading to unexpected or unintended behaviours.

This opacity is problematic, not least given that the home---a space in which may consumer IoT devices target---represents a private space (where such devices often feature sensors that can readily capture personal, sensitive and intimate information~\cite{crabtree2018}).
Furthermore, given the physical nature of such devices, one can readily envisage scenarios in which serious consequences might arise; smart ovens might automatically turn on during the night~\cite{carman2019}, or smart locks could prove obstructive if functioning unexpectedly, such as during a house fire~\cite{trimananda2020}.
As the IoT continues to pervade various aspects of our everyday lives, it is crucial that users are best placed to oversee, understand, and act upon any such issues, if and when they arise.

\subsection{The role of transparency in the consumer IoT} \label{subsec:transp_paradox}

\begin{figure}[!t]
\centering
\includegraphics[width=.9\columnwidth]{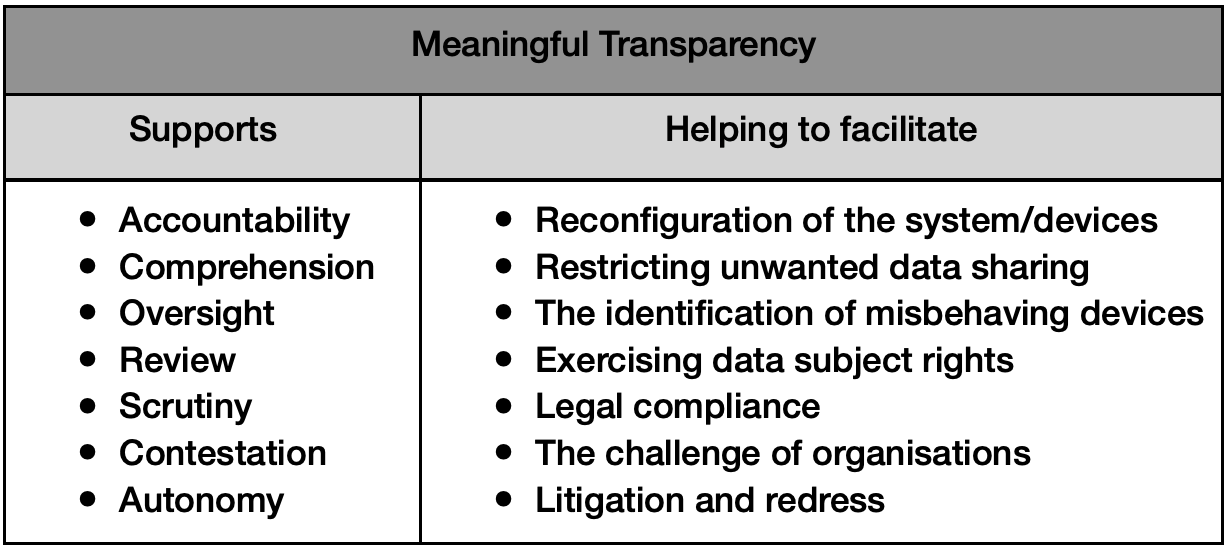}
\caption{Meaningful transparency supports stakeholders in understanding, taking action, contestation and enacting change.} 
\label{fig:intro:meaningful_transp}
\end{figure}

It follows that \textbf{transparency} in the context of the consumer IoT is important. 
Greater visibility over the consumer IoT can help users in a variety of ways, including satisfying curiosity and concerns about how things are behaving, ensuring that their data is appropriately handled, monitoring system behaviour, helping to ensure correct operation and functionality, revealing what lead to a particular fault or incident, and so on~\cite{singh2019}. 
That is, if users are able to better understand what is happening (or has happened) within their connected environment, then they are better positioned to take action and respond, as and when required.
Such actions might involve, for example, re-configuring parts of the system to prevent undesirable data sharing,  perhaps ceasing to use a particularly problematic IoT device altogether, seeking to contact a company to exercise data rights, begin litigation, and so on (see Fig.~\ref{fig:intro:meaningful_transp}).
Indeed, this {need for oversight} is increasingly recognised, as we see growing demands for transparency over technology, and emerging regulations requiring such (\S\ref{sec:disc:reg}).
Yet, the practical dimensions of transparency, \textit{as a means for supporting accountability}, is an under-considered area.

Importantly, simply providing information about a system is not necessarily useful (\cite{acquisti2013, bovens2007, obar2020, stohl2016, suzor2019,kamarinou2017machine}); providing \textit{too much} information, can even act to further obscure and obfuscate (the so-called `transparency paradox'~\cite{stohl2016}).
Rather, transparency must work to support users in understanding what's happening, so as to enable appropriate actions in response~\cite{bovens2007}.
In other words, through ensuring that users are able to understand and oversee the happenings of their smart environment (what data is being collected, how it is being processed, transmitted locally and/or shared remotely, why particular actions and outcomes are occurring, etc.), they can then take appropriate and informed decisions on what to do in reponse.

It follows that there is a need for \textbf{meaningful transparency}~\cite{suzor2019}: the provision of information in a manner that caters to the needs and expectations of the user to facilitate their effective oversight, scrutiny, and review over these technologies~\cite{cobbe2021}. 
In practice, such information should be \textit{contextually appropriate}~\cite{cobbe2021} to the user and their situation, and support wider aims of contestation~\cite{kaminski2021}, accountability~\cite{singh2019}, autonomy~\cite{pridmore2020}, and legal compliance~\cite{Ausloos2021} (see Fig.~\ref{fig:intro:meaningful_transp}).

In an IoT context, while there has been considerable focus on specific issues (such as those around security, privacy, complexity and supporting users with specific deployments), \textit{tackling opacity} so to support scrutiny and broader accountability aims has thus far been under-considered. 
Tackling such opacity in the consumer IoT is important, particularly as the prevalence of these devices increases~\cite{deloitte2021}, and as the consumer, social, and regulatory demands for greater transparency and accountability regarding technology continues to grow~\cite{pasquale2015}.

\subsection{Exploring meaningful transparency in the consumer IoT}

This paper explores the needs, expectations, and desires of participants as regards transparency mechanisms within the consumer IoT.
Given that we consider transparency for general consumers, and that many consumer IoT products aim at domestic use, our focus is on the \textit{smart home}. 
We consider the design of transparency mechanisms that are meaningful for their intended recipients (the users), exploring aspects such as what information (from, and about, their smart homes) they would want to know, how they believe this should be represented and communicated, and what such transparency mechanisms might look like in practice. 
In doing so, our goal is to provide practical ways forward by articulating types of \textit{user-derived requirements} (specifications that formalise the needs of users~\cite{holtzblatt1995}) and design approaches for those looking to build effective transparency mechanisms into the IoT.

To do this, we undertake three user studies, working with participants to explore the kinds of transparency mechanisms \textit{they} would want and expect from smart home systems. This is to reveal what they consider a meaningfully transparent smart home might entail, and provide tangible ways forward for researchers and designers alike.
\textbf{Study 1} takes a broader approach to develop an understanding of what users want to know (across a selection of scenarios where something within the smart home warrants attention).
We do this with a survey of 126 respondents, probing into what types of information they would care about, how they would expect that information to be communicated to them, and what follow-on actions they believe that this information would enable them to take.
To further explore this in practice, we conduct two co-design workshops (with 5--6 participants each) in \textbf{Study 2}.
During these sessions, the two groups of participants work together to create a list of user requirements and design prototype interfaces for what they would want and expect from a transparent smart home system.
The result of these workshops is i) a general set of design elements (i.e., key design features for enabling smart home transparency), and ii) a selection of categories representing the types of user requirements that our participants felt important for bringing about transparency.
\textbf{Study 3} then looks to generally test, validate and scrutinise these design elements and user requirement categories with a (new) set of 56 participants, indicating the generalisability of these findings.

\subsection{Contributions}

Exploring how transparency mechanisms can better support transparency, through enabling scrutiny and oversight, is an emerging area of general importance~\cite{norval2021, norval2022}. We consider this within the context of the consumer IoT. 
Towards this, we  offer practical ways forward for supporting {meaningful transparency} in smart homes, by providing:

i) categories of user-derived requirements to assist and guide developers in implementing transparency mechanisms within their offerings;

ii) a collection of co-designed transparency \textit{`design elements'} -- features of design for supporting smart home transparency; and 

iii) insights on some of the broader considerations and challenges facing the design of transparency mechanisms for bringing about a more accountable consumer IoT.

These contributions aim to better support users in gaining more effective oversight over their connected environments{, and developers and researchers in facilitating such}. 
Moreover, we argue the importance of transparency mechanisms that cater to the needs and expectations of their users, and discuss various practical ways forward (for the IoT, and, indeed, beyond) toward enabling more meaningfully transparent---and thereby supporting more accountable---technologies.

\section{Background}  \label{sec:background}

The \textit{Internet of Things} (IoT) typically refers to ``the extension of the Internet [...] into the physical realm, by means of the widespread deployment of spatially distributed devices with embedded identification, sensing and/or actuation capabilities''~\cite{miorandi2012}.
Importantly, however, there is much more to the IoT than just devices; 
a given IoT deployment will often comprise a socio-technical ecosystem of devices, software, systems, and organisations, in which the flow of data between, through, and across systems and organisations work to deliver overarching functionality~\cite{norval2021, singh2016}.
This may involve, for example, remote weather sensors (by means of web APIs), local sensors (proximity, temperature, etc.), and cloud-based web services, all working together to determine whether to actuate (turn on) a smart home's heating system.

IoT deployments raise the prospect of a range of potential issues occurring~\cite{singh2016}.
These might relate to system interoperability~\cite{trimananda2020}, a lack of manageability~\cite{brush2011, yarosh2017}, emergent behaviours~\cite{roca2016}, a lack of knowledge as to what data is being collected and how it is being used~\cite{jakobi2018, tabassum2019}, outright system failure~\cite{he2019, palekar2019, xing2021}, and so on.
However, the IoT entails opacity~\cite{norval2019}; it can be challenging for even the most technical users to effectively monitor, oversee, and/or diagnose how their smart environments are operating, be it to understand a particular issue or occurrence, or to satisfy curiosity. 
This can raise real challenges for the users of smart homes, with considerations relating to system reliability being said to play a crucial factor in smart home acceptance~\cite{park2017, schomakers2021}.

As such, there appear clear benefits for mechanisms that can support users with greater oversight and understanding over these complex and opaque systems; naturally, transparency can assist in this regard~\cite{obar2020, suzor2019}.
Transparency mechanisms can help illuminate the inner workings of a system -- they can support users in verifying that their system is operating in line with expectations, identify particular areas of interest, and conduct targeted investigations into specific components or behaviours that arise.
Moreover, effective transparency mechanisms also support individuals in making informed choices~\cite{crabtree2018}, allowing users to take further action, if and when required~\cite{norval2021}.
This might include the provision of information that results in users changing device settings to restrict undesired behaviours, removing problematic devices from use in the home, or challenging particular parties or contesting their actions~\cite{kaminski2021, yurrita2023} (such as through complaints, legal means, etc.). 
That is, \textit{effective transparency mechanisms will often be a precursor to pursuing accountability aims~\cite{singh2019,cobbe2021}}, playing a key role in supporting users in identifying, mitigating, and/or rectifying a range of different issues and concerns that may arise with regard to the consumer IoT.

\subsection{Specific user concerns} \label{subsec:attitudes}

In recent years, there have been numerous research studies exploring consumer attitudes to the IoT (see~\cite{marikyan2019, mocrii2018} for two systematic literature reviews).
Such research can provide valuable insights into the challenges and concerns facing prospective IoT users, across a range of different contexts and scenarios, and highlight areas where meaningful transparency may be able to assist.
As such, we now briefly discuss three particularly prominent concerns from across the literature.

\subsubsection{Understanding smart homes} \label{subsec:understanding_smart_homes}

Given that smart homes may be driven by a range of different systems (perhaps including various sensors, actuators, and online services, incorporating a range of organisational ecosystems)~\cite{singh2016}, there is real potential for IoT deployments to quickly become complex.
Even simple scenarios can lead to a disconnect between users' understanding of what is happening within their smart homes and what is actually occurring; 
in a user study with 20 participants, Yarosh and Zave~\cite{yarosh2017} {outlined a hypothetical smart lock (with four features of automated behaviour), and asked participants whether the door would be locked or unlocked within 20 given scenarios. 
They} found that participants' mental models did not appear to align with the operation of the system, despite the participants claiming to understand how these system interactions should work~\cite{yarosh2017}. 
This can be problematic, not least given the propensity for tech related issues within the home to be consequential (sensitive data may be captured and leaked, smart doors or windows may not secure when a person leaves the house, etc.).

Users misunderstanding what is occurring within their smart home has led to work which attempts to explore and mitigate this issue from a conceptual perspective.
For example, Despouys et al.~\cite{despouys2014} have proposed a model for `sensemaking' within the context of smart homes, in order to identify and manage potential scenarios where the expectations of users and autonomous systems are misaligned~\cite{despouys2014}.
Similarly, Chuang et al.~\cite{chuang2018} analysed `concept videos' of IoT products to develop a design vocabulary for human--IoT systems~\cite{chuang2018}, as a means for supporting developers in better communicating and explaining the intended purpose of their IoT systems, such that their users might have a stronger understanding of how they operate.
Some have explored the potential of methods for `explainable AI' within smart home contexts, given that consumer IoT systems will often employ some form of machine learning~\cite{dai2023, kok2023}.
Such works that support the comprehension of smart home systems offer one way forward for how the developers and designers of IoT systems might better manage and communicate the behaviours of their systems to their users, helping to ensure that such systems act in line with expectations.

Relatedly, other bodies of work have explored practical techniques or prototypes aimed at supporting users in understanding how their system is, or has been, operating.
Some of these have been more conceptual, such as work by Desjardins et al.~\cite{desjardins2020, desjardins2021}, who have used literary authors to translate IoT device data logs into fictional novels (`data epics') for their users~\cite{desjardins2020, desjardins2021}).
Others have {explored various techniques and modalities for presenting information about how smart devices are operating to their users,}
e.g. through providing descriptions of apps and devices~\cite{crabtree2018}, creating data visualisations for sensor feeds~\cite{castelli2017}, using `nutrition labels'~\cite{railean2018, railean2021}, designing interfaces for rule editors~\cite{manca2019, zhao2020, zhao2021}, using voice assistants to query smart home logs and provenance data~\cite{norval2019}, and recommendations to help users with issues of consent~\cite{castelluccia2018}.
The proposal of general mechanisms to help better inform users (as to the status and operation of their IoT devices) acts to highlight this topic as an ongoing area of concern.

It is clear that any potential disconnect between how users understand and expect their smart environments to operate, and how those environments may actually function can have negative implications.
However, while some of the above work offers empirical insights into how we might better support users in understanding how their smart environments operate, much of the research into this topic has thus far involved participants only at the evaluation stage of developing a particular proposed solution or approach. 
Less considered are the needs and expectations from the perspective of users themselves -- \textit{what they want to know}, \textit{how they believe that this should be communicated to them}, etc.
Engaging users, however, as part of a design-process is important for helping ensure that its outcomes are helpful and effective~\cite{abras2004user, yao2023}.
Therefore there are clear opportunities for research that is more formative in nature, focusing on how best we can support potential users---taking into account their own perspectives---in understanding how their IoT systems operate.

\subsubsection{Privacy}  \label{subsec:privacy}

Another specific issue for the IoT relates to user privacy; privacy is undoubtedly an important consideration within a consumer IoT context, where concerns include who data is being shared with~\cite{emaminaeini2019, haney2020, haney2021, kulyk2020, marky2020a, williams2017, windl2023, zheng2018}, how that data is being used by organisations~\cite{abdi2019, chalhoub2021, emaminaeini2019, haney2020, haney2021}, the intrusiveness of device sensors (e.g. cameras, microphones)~\cite{chalhoub2021, emaminaeini2019, haney2020, haney2021, zeng2017}, and broader concerns over a loss of control of personal information~\cite{chalhoub2021, emaminaeini2019, haney2020, haney2021, kulyk2020, zheng2018}.
Indeed, such concerns are for good reason, with recent studies showing that a considerable number of consumer-IoT devices are `leaky', in the sense that they involve substantial, potentially non-essential, communication with third-parties~\cite{hudig2023, mandalari2021, ren2019}.
Further exacerbating these issues is that consumer IoT devices will often be used within the home, where devices can potentially capture personal and intimate information~\cite{crabtree2018}, and `privacy norms' may easily be violated~\cite{apthorpe2018}.
As a result, such privacy-related concerns have been identified as key factors affecting users' purchase behaviour~\cite{emaminaeini2019, marky2020a}, and whether they trust~\cite{kulyk2020, liao_understanding_2019}, accept~\cite{lafontaine2021, schomakers2021}, and/or adopt~\cite{liao_understanding_2019} smart devices within their homes.

While privacy remains an ongoing challenge in the IoT, there have been a plethora of research efforts that show promise in tackling particular privacy issues. 
For example, there have been a number of recent developments with regard to privacy enhancing technologies for the IoT (see~\cite{cha2019, li2019}), as well as researchers proposing methods for the developers of domestic IoT devices to better meet the privacy expectations of users, for example, by way of privacy norms~\cite{abdi2021}, privacy-oriented design implications~\cite{yao2019c, yao2019b}, and recommendations~\cite{chalhoub2020a, chhetri2019b, haney2021, zeng2019, zheng2018}.
Researchers have also developed and evaluated storyboards~\cite{jin2022}, prototypes~\cite{chhetri2022}, and tools~\cite{seymour2020} for assisting privacy-oriented concerns, privacy controls which provide varying levels of detail about why certain data is being requested~\cite{marky2020b}, and nutrition labels~\cite{emaminaeini2019} to inform consumers across a range of IoT-related privacy issues.

Importantly, while many of these concerns have (understandably) resulted in privacy-oriented outputs (be they interventions, prototypes, designs, resources, etc.), there is also scope for broader transparency mechanisms to assist here.
That is, transparency mechanisms can support greater levels of oversight and understanding over smart home operations, and this information can, in turn, assist in privacy contexts.
For example, such information might reveal the data being collected by an IoT device, where that information flows, how it is being used, and so on, which can indicate potential privacy concerns.
Furthermore, such mechanisms also enable various actions to be taken in response, perhaps prompting users to change privacy settings, restrict where data is flowing to (e.g. see~\cite{mandalari2021}), engage with particular parties to challenge or exercise data protection subject rights (as has been discussed at a high-level~\cite{castelluccia2018, hudig2023, seymour2020, urquhart2020, wachter2018}), and so on.
Once again, research which focuses on the transparency needs and expectations of users will have much to offer, and complement privacy-oriented research into the IoT.

\subsubsection{Security} \label{subsec:security}

The IoT also gives rise to a range of security issues~\cite{komninos2014, meneghello2019, mohanta2021, sabir2022, singh2016, urquhart2020, zhao2013}.
IoT deployments may comprise a large number of different devices (perhaps manufactured by a range of organisations, with some having more stringent security practices than others), and thus potentially introducing numerous possible points of failure~\cite{norval2021}.
Moreover, the IoT will often entail physical elements (e.g. actuations), where the consequences of security incidents may be severe by resulting in physical world harms~\cite{carman2019, trimananda2020, singh2016}.
And again, IoT devices can reside within intimate or sensitive locations within the home (e.g. bedrooms, bathrooms)~\cite{crabtree2018}, making the prospect of security issues particularly concerning.
In all, security issues within the consumer IoT is an ongoing concern.

Unsurprisingly, IoT security is  a notable concern for consumers; work exploring consumer attitudes to IoT adoption have identified a number of particular reasons for this, including concerns over data breaches~\cite{chalhoub2021, haney2020, haney2021}, the perceived challenges of achieving a secure smart home~\cite{brush2011, chalhoub2020b, haney2021, hwang2015}, access and power imbalances in multi-user smart homes~\cite{geeng2019, zeng2019}, the intrusiveness of another party being able to access sensor data (e.g. cameras, videos)~\cite{chalhoub2021, duezguen2021, emaminaeini2019, haney2020}, the risk of having physical devices manipulated by a malicious actor (e.g. switching off the freezer, manipulating smoke detectors~\cite{chalhoub2021, duezguen2021}), and the perceived risk of physical safety (e.g. unlocking doors or windows)~\cite{duezguen2021, haney2020, haney2021}, among others.

Again, mechanisms for improving IoT transparency may have a role to play in relation to many of the above security concerns, for example, through providing information and oversight over the operation of devices (i.e. what drove particular outcomes and where data is flowing to~\cite{singh2019}), to help confirm that devices are interacting and behaving in line with expectations~\cite{trimananda2020, pasquier2018}, and in identifying anomalous behaviours and possible security threats~\cite{han2020ndss, wang2021}.
Such information (be they access logs, provenance information~\cite{singh2019,pasquier2017}, reports, or other forms of ongoing system monitoring~\cite{singh2018, pasquier2018}) can facilitate general oversight, targeted investigations, and subsequent follow-up actions to mitigate these concerns -- perhaps prompting the user to seek further information from device manufacturers, or even removing components from the IoT deployment altogether~\cite{norval2021}.
We are already seeing some proposed solutions for communicating such security-related information to users (again, such as nutrition labels~\cite{emaminaeini2019, emaminaeini2020}).

\subsection{Towards meaningful transparency: Addressing a gap} \label{sec:meaningful_transparency}

Importantly, in each of these three concerns, effective transparency mechanisms have much to offer, through better supporting users in overseeing and monitoring how their smart homes are operating, helping to identify and diagnose potential issues as they arise, and enabling informed responses as a result. 
Specifically, this can help mitigate issues of \textit{understandability} (\S\ref{subsec:understanding_smart_homes}), through providing relevant information about the system's operation in a way that is meaningful to the user, thereby supporting them in comprehending why the system is behaving in the way that it is.
Such information can also help address \textit{privacy} concerns (\S\ref{subsec:privacy}), through providing clarity over when and where data is being transferred, which can facilitate follow-on actions to prevent, mitigate or seek recourse for issues relating to privacy. 
Finally, these oversight mechanisms can also help issues relating to \textit{security} (\S\ref{subsec:security}), through supporting users in verifying their correct operation (thereby assuaging such concerns), and in identifying, investigating, and seeking support and restitution in the event of security breaches.
In all, meaningful transparency mechanisms allow users to monitor for, uncover, and act upon many of the types of concerns raised in the literature (as just discussed).

However, the question of how transparency mechanisms can best support users, such that they are meaningful and effective, is currently under-considered. 
There is therefore a clear role for research towards more effective transparency mechanisms by \textit{working with potential users throughout the design process} to explore how transparency mechanisms can better reflect the needs and expectations of those that would stand to benefit from them.
This means that there are research opportunities towards advancing our understanding of what meaningful transparency mechanisms in the consumer IoT might actually entail (as is our focus)---focusing on what users want to know, how they expect to interact with such, the benefits they perceive such mechanisms would provide, and so on---and to explore how such mechanisms might support wider concerns of understandability, privacy, and security that arise in the literature more broadly.

To reiterate, the topic of smart home transparency is important -- through providing information over the operation of smart home systems, transparency mechanisms help support various follow-on actions in response to particular issues.
Having greater levels of oversight might prompt users to change system settings to prevent egregious behaviours, verify correct system operation, remove devices from the IoT deployment entirely, or even identify and challenge particular parties (such as through legal means), etc. 
That is to say, having transparency mechanisms that are more effective, by being user-derived and user-centric, can work to assist users with a  range of different challenges, issues, and concerns that they may face regarding the IoT.

\section{Methodological Overview} \label{sec:method}

Our approach entails working with prospective IoT users to  explore, design, and evaluate mechanisms for bringing about greater levels of transparency within the consumer IoT.
Specifically, we explore the  general needs and expectations that users have with regard to transparency mechanisms, collect and categorise transparency-related user requirements, and develop and identify elements of design that the participants felt would enable meaningful oversight within a selection of scenarios.
To do this, we undertake three user studies:

\textbf{Study 1} (\S\ref{sec:survey}) involves a survey to explore what transparency information they would consider useful within a smart home context, how they feel that information should be communicated to them, and how they consider one might use that information (i.e. what they foresee such information enabling).

Study 2 (\S\ref{sec:focusgroups}) comprises two independent workshops of 5--6 participants, representing the `users' of smart homes.
During each workshop, the participants work together to identify user requirements and design system prototypes that represent what they believe to be effective transparency mechanisms for the consumer IoT.
This is to gain detailed insight into the types of transparency mechanisms that participants thought would be useful to realise, and how they foresaw using such interventions in response to particular interests, concerns, or events.

\textbf{Study 3} (\S\ref{sec:eval}) takes the insights from the prior two studies, and validates their relevance and applicability through an evaluation survey with an altogether new set of 56 participants.
Here, we look to explore the efficacy of our previous findings with a new sample, and find that our results from Study 2 appear to generalise (to this new cohort).

Our methodology was structured such that each study builds upon the findings and insights of those prior, while uncovering related insights and discussion points throughout the process.
Note that all of our studies were approved by our departmental ethical review board, and all participants were compensated for their time in an amount reflecting the UK's `living wage'.

\subsubsection{Participant recruitment} \label{subsec:methodology_recruitment}

For our surveys (Studies 1 \& 3), we used Mechanical Turk (MTurk) for participant recruitment. 
MTurk is a widely popular recruitment method within academic research~\cite{pew2016}, and allows us to purposefully `cast the net wide', to document a diverse set of different perspectives and ideas that we could find, enabling sampling at scale.
It is for this reason we placed few restrictions on demographics so as to enable a range of participants, though we did enact some restrictions.
For example, we restricted to those likely to have some command of English, and have a good MTurk task success rate (in line with guidance from other studies~\cite{shipman2020}).
We also, purposefully, did not restrict the survey to IoT users (so as not to exclude perspectives of those who have not yet adopted such devices), nor did we limit participants based on levels of technical expertise (so to provide insights from various backgrounds).

For the co-design workshop (Study 2), we selected two groups of participants; 
the first group comprised respondents of Survey 1, those with a general background bringing with them a broad set of perspectives.
However, given that Study 2 involves creating user requirements and involves visual design, we recognised that there were also advantages in recruiting participants with some level of experience in technology design, so as to provide a complementing and contrasting perspective to that of the other group.
For this we recruited a second group comprising a `convenience sample'~\cite{jager2017} of undergraduate computer science students.
This allowed us to leverage their experience of user experience (UX) principles, and offered a point of comparison (to explore the similarities and differences) between the students and the more `general' users.
Full details of participant recruitment are described with each study, and we further discuss the implications of our participant samples in \S\ref{subsec:limitations}.

\subsubsection{Grounding the studies} \label{subsec:methodology_grounding}

Across all studies, we make use of a selection of scenarios to ground and motivate our work.
Each of these are carefully designed to reflect actual concerns that people have, with many motivated by examples discussed in the literature and observed in the real-world, including issues associated with data leakage~\cite{chalhoub2020c, mandalari2021, winder2020, yus2022}, physical harms~\cite{trimananda2020}, system malfunction~\cite{he2019, palekar2019, xing2021}, targeted advertising~\cite{winder2020}, system operation~\cite{brush2011, carman2019, norval2019, roca2016, trimananda2020, yarosh2017}, etc.
Further information about each study are elaborated in their respective sections, and all data has been made available via GitHub~\cite{norval2023git}.

\section{Study 1: Surveying user interests and expectations for transparent smart homes} \label{sec:survey}

Our first study entailed an online survey  designed to uncover a broad understanding of what meaningful transparency within a consumer IoT context might entail.
We focus on four main questions; given a scenario where something within a smart home warrants attention or goes wrong: i) what types of information do respondents feel is important to know?; ii) how do they expect this information to be communicated (in terms of system interaction)?; iii) how do they expect this information to be presented (in terms of design)?; and iv) what types of follow-on actions do they think this information would enable them to take?
By exploring these questions, we obtain a better understanding (particularly within a smart home context) of the types of transparency mechanisms that users might come to expect, how these might work to meaningfully communicate the relevant information to the user, and how our participants would seek to use such mechanisms to support wider accountability aims.

\subsection{Method} \label{sec:survey:questionnaire}

We recruited 126 participants from Mechanical Turk to take part in this questionnaire (survey). 
As Table~\ref{tab:demo1} shows, these respondents reported a range of technical expertise: 23\% claimed to have `no' or `some' knowledge; 32\% had `average' knowledge, and 45\% had `advanced' or `expert' knowledge, thus reflecting various aims, understandings, and expectations.
While we did not restrict the survey to only those actively using consumer IoT devices (\S\ref{subsec:methodology_recruitment}), we nevertheless observed that 94\% reported having smart devices within their homes, demonstrating that the vast majority of our respondents were actively being impacted by consumer IoT devices, and thus had some familiarity as to what the IoT represents.
In this way, our sample is reflective of the aims earlier mentioned (\S\ref{subsec:methodology_recruitment}) comprising a range of interests, expectations and end-goals, while still having a baseline understanding of consumer IoT products.

\begin{table}[h]
    \centering
    \caption{Demographic information of the participants from Study 1.}
    \label{tab:demo1}
    \scriptsize
    \begin{tabular}{ l l }
    \hline
    ~ & \textbf{\% of respondents}\\
    \hline
     Gender & ~ \\ 
     \quad Female & 31\% \\  
     \quad Male & 69\% \\
     \quad Other & 0\% \\
     \hline
     Age & ~ \\ 
     \quad 18--29 & 35\% \\  
     \quad 30--39 & 37\% \\
     \quad 40--49 & 20\% \\
     \quad 50--59 & 8\% \\
     \quad 60+ & 0\% \\
     \hline
     Technical Expertise & ~ \\ 
     \quad No knowledge & 5\% \\  
     \quad Some knowledge & 18\% \\
     \quad Average level of knowledge & 32\% \\
     \quad Advanced knowledge & 34\% \\
     \quad Expert knowledge & 11\% \\
     \hline
     Knowledge of Smart Devices & ~ \\ 
     \quad No knowledge & 1\% \\  
     \quad Some knowledge & 25\% \\
     \quad Average level of knowledge & 53\% \\
     \quad Advanced knowledge & 39\% \\
     \quad Expert knowledge & 8\% \\
     \hline
     Have Smart Devices in the Home & ~ \\ 
     \quad None & 6\% \\  
     \quad One or more & 94\% \\
     
     \hline
    \end{tabular}
\end{table}

\begin{table*}[t]
  \caption{The four scenarios that participants were asked about.}
  \label{tab:briefdesc}
  \footnotesize
  \begin{tabularx}{\textwidth}{cXp{6.5cm}}
  
    \hline
    \textbf{Scenario} & \textbf{Brief Description} & \textbf{Underlying Information}\\
    \hline
    
    Voice Assistant & 
    You have seen news reports about certain smart voice assistants constantly recording audio and sending it to the manufacturer. 
    Given this, you wish to check your own smart assistant to verify what, and when, information has been sent outside of your home. &
    After further investigation, you learn that i) information is only being sent to the voice assistant's manufacturer when it specifically hears its trigger word (its name being called); ii) an audio recording is then sent to the manufacturer for processing, and an audio response is sent back; iii) the voice assistant is seen to be communicating with a number of different advertising companies. \\ \hline 

    Smart Windows &
    You wake up in the middle of a particularly cold night, noticing that your smart windows have opened automatically. You know that the windows are automatically set to open when it is above a certain temperature inside. However, the indoor temperature feels far too cold for this occur, and the windows should therefore not have opened. &
    After further investigation, you know that your smart home contains three indoor temperature sensors, which are accessed by the smart home system. These readings are used to determine whether or not the windows should be opened (as well as for other purposes).
    One of these devices has been reporting temperatures far higher than the other two, suggesting it may have malfunctioned. These unusual readings began just after midnight. \\ \hline
    
    Smart Fridge &
    Your Smart Fridge allows you to keep an inventory of what is stored inside, and can build a shopping list for use on-the-go. You receive an email from the fridge's manufacturer, updating the terms of service to allow this data to be used for advertising purposes with other companies. As a result, you are concerned about the privacy implications of your shopping habits being used for advertising. &
    After further investigation, you find that your fridge is categorising the types of items that you buy in order to predict characteristics about you (e.g. `vegetarian', `health conscious'), sending these the manufacturer.
    Further, you learn that the fridge is also sending this information to supermarkets and other advertisers, allowing them to send you adverts which they think will be of interest. \\ \hline

    Smart Locks &
    You have a smart lock which should automatically lock your front door every evening after sunset. One evening, you notice that something is preventing the door from locking, despite it being dark outside. &
    Your smart home works to unlock the front door whenever you arrive home. This works by a sensor that detects when your mobile phone is near the smart lock.
    Investigating the issue, you discover that your mobile phone is still being detected by this sensor, despite it being several rooms away from the front door. This appears to be preventing the front door from locking. \\

  \hline
\end{tabularx}
\end{table*}

After signing up and providing consent to participate, the main body of the questionnaire posed a selection of hypothetical scenarios to the participants, who were then asked open-ended questions relating to the above four questions.
These scenarios, outlined in Table~\ref{tab:briefdesc}, were based on real-world concerns and incidents; one related to suspected data leakage from voice assistants~\cite{chalhoub2020c, winder2020}; one concerned smart windows opening when they shouldn't have~\cite{alavi2019}; one related to targeted advertising within a smart fridge~\cite{bastos2018, yus2022}; and one involved a smart lock not behaving as expected~\cite{trimananda2020}.
In this way, our scenarios were grounded in actual issues that users of such devices might face.

Participants were initially presented a \textit{brief description} of one of the four scenarios (randomly selected), where some particular issue or concern warranted attention (see Table~\ref{tab:briefdesc}).
These, again derived from real-world scenarios (\S\ref{subsec:methodology_grounding}), were written in such a way as to have multiple possible reasons for the issue occurring, and the exact nature of what we had determined was happening within the smart home was withheld to participants at this stage of the survey.
Participants were first asked what they thought was the most important information to know in that scenario. 
They were then taken to a new page, containing some further \textit{underlying information} about the exact nature of that scenario, and were then asked: how should this information be communicated to the user; how might this information be structured or presented; and what follow-up actions would having access to such information facilitate.
This process was then repeated with a second scenario (randomised ordering), giving us a range of responses across all of the scenarios while limiting the time commitment required of the participants.\footnote{Note that as part of a pilot run, a few of the participants did complete this process for all four scenarios, before the number of scenarios presented to each participant was reduced to two to better reflect the anticipated time commitment that the study required.}

\begin{figure*}[!t]
\centering
\includegraphics[width=.85\textwidth]{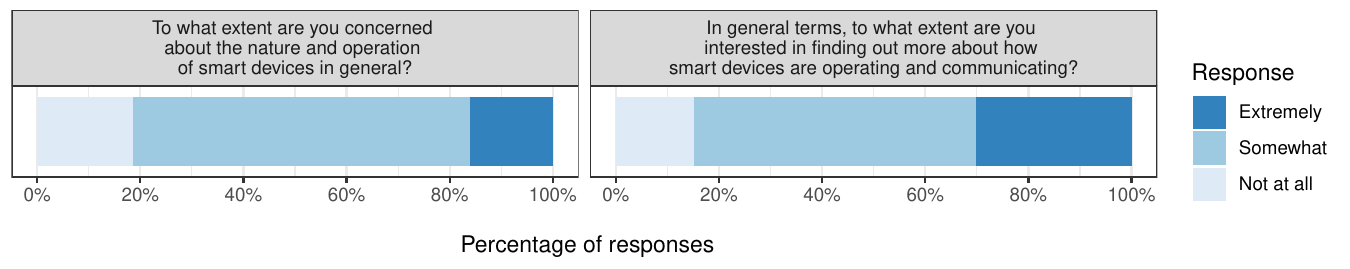}
\caption{The majority of Study 1 respondents were concerned about the nature and operation of smart devices, and interested in finding out more about them.} 
\label{fig:survey:concerned}
\end{figure*}

After completing the two scenarios, participants were then asked some broader (general and demographic) questions about themselves and their attitudes to smart homes.
These included questions relating to themselves {and} their technical expertise (Table~\ref{tab:demo1}), and their broader concerns and interests relating to smart homes more generally (Fig.~\ref{fig:survey:concerned}). 
Finally, we included an optional opt-in field where participants could enter their email address if they were happy to be contacted for subsequent stages of the research. 
All open-ended questions were analysed using thematic analysis~\cite{braun2006};
answers (either in part, or in entirety) were categorised under multiple themes, and this was conducted iteratively until themes no longer changed as a result of new data.

\subsection{Findings} \label{sec:survey:findings}

Our initial survey uncovered a range of insights about meaningful transparency within a consumer IoT context.

\subsubsection{Most respondents had concerns  about the nature of smart devices, and wanted to know more about their operation}

Asked directly, we found that over 80\% of respondents were at least somewhat or extremely concerned about the nature and operation of smart devices in general (see Fig.~\ref{fig:survey:concerned}). 
Of those who selected ``I'm extremely concerned'', further probing on their open-text elaborations for this answer revealed that this was largely either due to surveillance and privacy concerns, or device security and personal safety concerns (each of these categories making up $\sim$1/3 of ``extremely concerned'' responses).
This corresponds with the topics generally focused on in the literature (\S\ref{subsec:attitudes}), while clearly indicating the importance of transparency mechanisms to help support such issues. 
In contrast, of the 19\% that were ``not at all'' concerned, this was typically due to them not believing that ``there is reason to be concerned'', with $\sim$1/3 of these responses elaborating the sentiment that ``there isn't any useful information someone is going to get out of my [data]''. 
The remaining group of ``somewhat concerned'' respondents seemed open to the idea of using smart devices in the home, but still expressed caution over how they operated. 
Again, when asked directly, over 80\% of respondents indicated that they were either somewhat or extremely interested in finding out more generally about how smart devices were operating and communicating (see Fig.~\ref{fig:survey:concerned}). 
Such findings demonstrate the appetite for greater transparency, and for having the ability to oversee, inspect, and understand what is happening within smart homes.

\subsubsection{Many respondents wanted to be able to oversee technical specifics}

\begin{table}[h]
    \centering
    \caption{A thematic analysis from Study 1. Themes were coded from participants' open-text responses, indicating what information they felt was important to know for a given scenario.}
    \label{tab:survey1_coding}
    \scriptsize
    \begin{tabular}{ l r }
    \hline
    \textbf{\% of responses categorised under each theme} & ~\\
    \hline
    
     \textbf{Voice assistant} (n = 67) & ~ \\ 
     \quad What information was recorded by the voice assistant & 38.8\% \\
     \quad Why the information was recorded by the voice assistant & 16.4\% \\
     \quad When the information was recorded by the voice assistant & 10.4\% \\
     \quad Where the information was stored & 6.0\% \\ 
     \quad What information was transferred over the network & 52.2\% \\
     \quad When information was transferred over the network & 19.4\% \\
     \quad Why information was transferred over the network & 6.0\% \\
     \quad How long information will be stored by other parties & 6.0\% \\
     \quad Where information will be stored by other parties & 10.4\% \\
     \quad Who has access to the information once sent to other parties & 20.9\% \\
     \quad How information is used by other parties & 22.4\% \\
     \quad How to prevent this from happening in the future & 25.4\% \\
     \hline
     
     \textbf{Smart Windows} (n = 74) & ~ \\ 
     \quad Why the window opened & 48.6\% \\
     \quad When the window & 12.2\% \\
     \quad What data source triggered the actuation & 48.6\% \\ 
     \quad What data (readings) drove the actuation & 47.3\% \\
     \quad How to prevent this from happening in the future & 14.9\% \\
     \hline
     
     \textbf{Smart Fridge} (n = 69) & ~ \\ 
     \quad What information was recorded by the smart fridge & 7.2\% \\
     \quad What information was transferred over the network & 53.6\% \\
     \quad When information was transferred over the network & 8.7\% \\
     \quad Where information will be stored by other parties & 5.8\% \\
     \quad Who has access to the information once sent to other parties & 34.8\% \\
     \quad How information is used by other parties & 44.9\% \\
     \quad How to prevent this from happening in the future & 33.3\% \\
     \hline
     
     \textbf{Smart Locks} (n = 62) & ~ \\ 
     \quad What happened to the smart lock & 3.2\% \\
     \quad Why the smart lock didn't secure & 35.5\% \\
     \quad What data source prevented the door from locking & 45.2\% \\ 
     \quad What data (readings) prevented the door from locking & 25.8\% \\
     \quad How to prevent this from happening in the future & 43.5\% \\
     
     \hline
    \end{tabular}
\end{table}

We again used thematic analysis~\cite{braun2006} to explore what participants thought was important to know within each particular scenario. 
Open-text responses were categorised, and themes raised by at least two participants are presented in Table~\ref{tab:survey1_coding}, alongside their prevalence (the proportion of responses that were categorised under each theme). 
Note that given participants were only shown two of the four scenarios (randomised), each scenario has a different number of responses.
The full set of responses and their codes are included within the supplementary materials.

One of the main themes to emerge, particularly for the window and smart lock scenarios, was the need to understand and verify what was going on. 
Indeed, nearly half of the responses for these two scenarios wanted information which allowed the participant to understand and verify what drove this actuation (49\% and 45\% respectively), such as accessing ``a history log'' of the smart home, ``to see why this triggered'' and ``whose fault it was''.
This is particularly interesting, given the literature on understanding smart homes (\S\ref{subsec:attitudes}), where there was a potential disconnect between users' understanding of what was happening within smart homes and what was actually going on~\cite{yarosh2017}.

Looking to the voice assistant and smart fridge scenarios, there was also an appetite for overseeing technical specifics -- though this time predominantly regarding data flows. 
For example, the most frequent theme emerging from these responses indicated the importance of information regarding \textit{what} was being transferred outside of the smart home (52\% and 54\% of responses for the voice assistant and smart fridge scenarios respectively), \textit{who} could access that information (21\% and 35\%), and \textit{how} that information was being used (25\% and 33\%). 
However, these two scenarios also tended to raise the topic of \textit{prevention} for this unexpected smart home behaviour; over a quarter of responses for the voice assistant and a third of responses for the smart fridge scenarios specifically indicated that it was important to be able to ``opt out'', ``leave this service", and ``stop this level of invasive behaviour''.

\subsubsection{Interaction techniques and information visualisations offer one way forward}

A common suggestion raised by participants with regard to how they expected this information to be conveyed to them was the desire for automatic notifications when a discrepancy or some anomaly was detected (such as the example of the faulty sensor providing readings far higher than would be normal). 
For example, one commented ``[I'd want] to be automatically notified if a discrepancy between them is recorded at any point''. 
This was typically described by respondents as an alert on the system, a mobile notification, an email, a phone call, etc.
There were also descriptions of means to allow the user to perform more targeted investigations (i.e. in response to a particular incident or concern), alongside those driven by interest and more exploratory in nature.  
In terms of how that information should be communicated, responses were often contextual.
For example, in the voice assistant scenario, some referred to asking the voice assistant, as a means to interrogate what is going on -- a method previously suggested in the literature~\cite{norval2019}.
In contrast, for the window scenario, many of the responses described graphs of temperature readings (which have also been previously explored~\cite{castelli2017}).
Other suggested responses included lists and tables (of raw numbers), summarised explanations, schematic diagrams, and other forms of data visualisation.

\subsubsection{Enabling a means to take control} \label{subsec:survey:control} 

On the subject of what types of follow-on actions such information would enable, a few common themes that emerged related to the ability to disengage with the device or devices in question, should they be behaving in undesirable ways. 
For example, for the scenarios concerning user privacy (the voice assistant and smart fridge), respondents indicated that they would reassess their use of the device in question some sample quote include: ``I would restrict my use of this voice assistant'', ``I'd have to seriously consider if owning a smart fridge was right for me'', ``this is a smart device that I just wouldn't have''.
Again, this demonstrates the key role that transparency information can play, for example as regards technology acceptance and adoption, by showing how it enables and supports effective oversight and actions in response to what occurs.

\subsection{Summary}

Throughout this survey, we probed respondents for their thoughts, interests, needs, and expectations as regards transparency mechanisms for the consumer IoT.
From the results,   we can see that participants did express concerns about how smart devices were operating, and there appeared real appetite for transparency measures that granted access to such information -- both when things began to go wrong, and for wider aims of curiosity, validation, etc. 
The results indicate that there is much scope for research that explores how such transparency mechanisms might come to be expected by potential smart home users in practice.
In all, the respondents appeared to recognise the benefits of having the ability to oversee their smart homes, and the many opportunities of doing so.

\begin{figure*}[!t]
\centering
\includegraphics[width=\textwidth]{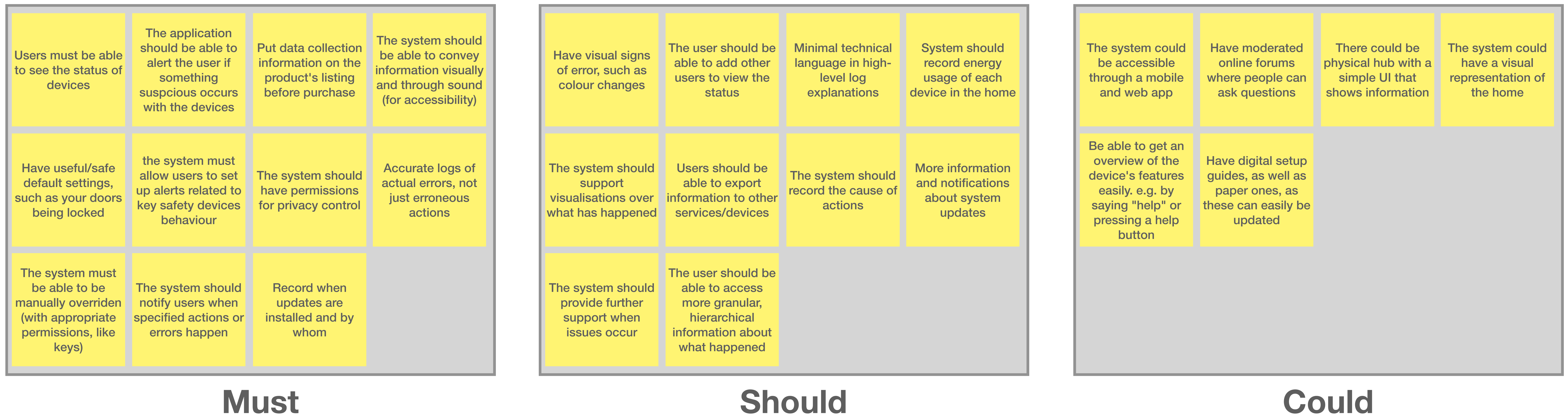}
\caption{A screenshot from Group 2's MURAL board. Participants could co-create requirements (sticky notes) and prioritise them as either `Must', `Should', or `Could' by moving them into boxes.} 
\label{fig:fg:board}
\end{figure*}

\section{Study 2: Co-designing transparency requirements and design elements} \label{sec:focusgroups}

In the previous study, participants had clear concerns over the nature and operation of smart devices, and overwhelmingly expressed an interest in having effective transparency mechanisms for the consumer IoT. Building on this, we next further probed on how IoT users believe that such transparency mechanisms should work in practice. 
To do this, we take a user-centric approach~\cite{abras2004user}, undertaking two co-design workshops with participants to derive \textit{their key considerations and design ideas} for transparent smart home systems.
By doing so, we uncover the types of transparency mechanisms that our participants want and expect from the consumer IoT.

\subsection{Method} \label{sec:focusgroups:method}

Our two workshops entailed the same process, each lasting two hours and involving separate groups of participants.
These participants were tasked with working together to complete two activities.
The first activity involved the participants deriving user requirements for making smart homes more meaningfully transparent, allowing us to explore the types (or categories) of user requirements that they thought were important. The second activity involved the participants designing visual prototypes for investigating and understanding the operation of smart homes, which allowed us to derive a series of \textit{design elements} (key aspects and features of design for enabling smart home transparency). 
These two activities were designed to provide tangible insights and ways forward---from the perspective of potential users---for how we might bring about consumer IoT systems with meaningful transparency in mind.

Our two groups were selected to bring a range of complementary skillsets and different perspectives, by having some with grounded experience in design vs. more `general' users.
For the latter, we began emailing all participants from the prior survey (\S\ref{sec:survey}) that indicated that they would be interested in taking part in follow-up research.
This email outlined the nature of the follow-up research, and asked those interested to complete a Doodle poll indicating dates that they were available.
This resulted in six participants that were all available for a particular time slot (though one of these participants ultimately did not attend).
This group therefore represented a fairly general group of prospective users (though with all having at least one IoT device within their homes).

Our second group comprised individuals with some knowledge of system requirements and interface prototyping, as a means of ensuring that their responses were grounded within some degree of systems design.
To achieve this, we sent an email through our institution's computer science department (i.e. a `convenience sample'~\cite{bornstein2013, jager2017, Sedgwick2013, stratton2021}), and recruited six undergraduate students to take part in this workshop.
In addition to their knowledge of systems design, this cohort also allowed us to compare and contrast their outputs to that of the first (more `general' or `standard' user) group, to explore where similarities and differences might exist, and what insights might be learned as a result.

\begin{figure*}[!t]
\centering
\includegraphics[width=\textwidth]{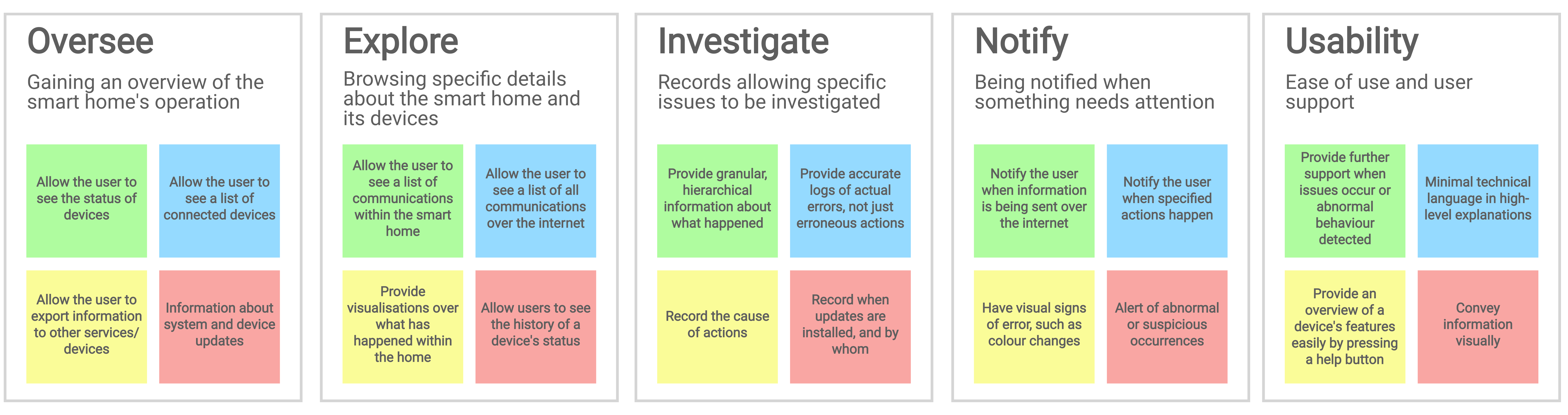}
\caption{Derived categories of requirements (with examples) that focused on transparency.}
\label{fig:fg:transparency}
\end{figure*}

\begin{figure*}[!t]
\centering
\includegraphics[width=0.9\textwidth]{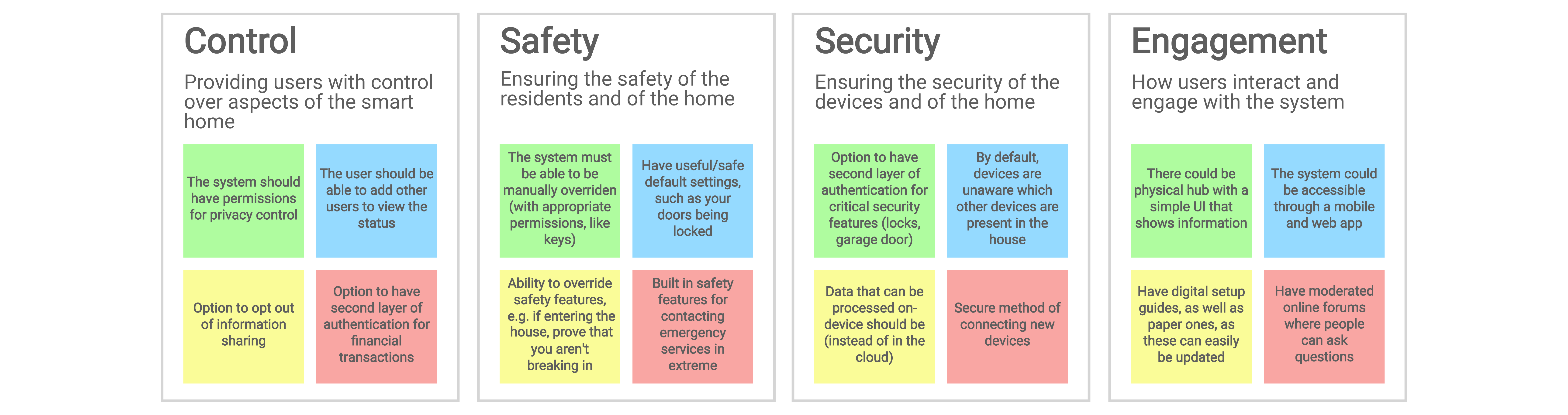}
\caption{Other categories of derived requirements (with examples) regarding broader concerns.}
\label{fig:fg:wider_reqs}
\end{figure*}

\subsection{Activity 1: Co-designing transparency requirements}

\label{sec:fg:reqs} Participants were given an overview of the research, including what requirements are, what constitutes a `good' requirement, and how they can be created.
They were then asked to consider how smart homes might better cater to their interests and concerns, and to think about how these might be specified as user requirements.
The activity itself was done on MURAL~\cite{mural2022} (a collaborative whiteboard web app), and involved participants creating requirements (sticky notes with text), as well as moving and editing those created by themselves or others. 
Participants were also given access to some example quotes from the prior study as a means to stimulate discussion, though they were encouraged to include and contribute any of their own requirements should they have some in mind.
Throughout this $\sim$20 minute exercise, the participants were tasked with determining user requirements and prototype functionality as they saw fit.
During this time, the researcher acted in a supportive role, chairing the co-design sessions careful to avoid biasing the outcomes with the researchers' preconceptions~\cite{smith2014}.
Once complete, participants then prioritised their requirements into three categories; `Must', `Should', and `Could' (see Fig.~\ref{fig:fg:board}), in line with the MoSCoW method of requirements prioritisation~\cite{elsood2014}.

\subsubsection{Activity 1 Findings -- Categories of transparency requirements} \label{sec:fg:req:findings}

This first activity led to 55 requirements being created across the two groups, with Group 1 creating 28, and Group 2 creating 27.
To gain a better understanding of the types of requirements that were produced, each requirement was coded using thematic analysis~\cite{braun2006} (in line with the process outlined in \S\ref{sec:survey:questionnaire}) after the workshops had taken place.
This was done so that we could explore not only the exact requirements that were identified, but the broader patterns and concerns that the participants focused upon. 
Through this process, a total of nine categories were identified; five comprised transparency-related concerns (Fig.~\ref{fig:fg:transparency}), the remaining four concerned broader controls and mechanisms that the smart home should support (Fig.~\ref{fig:fg:wider_reqs}).
These categories offer various insights into the types of transparency concerns that users may have, and the types of user requirements that may address these.

Given the focus of this work (on understanding how participants themselves felt transparency mechanisms should be achieved), our particular interest is in the five requirements presented in Fig.~\ref{fig:fg:transparency} (in contrast to those requirements concerning broader controls and mechanisms that the system should enable; Fig.~\ref{fig:fg:wider_reqs}).
These five transparency requirements focus on \textit{overseeing}, \textit{exploring}, and \textit{investigating} aspects of the smart home, as well as being \textit{notified} when unusual behaviours or activities are identified, and conveying the relevant information in a \textit{usable} way.
As shown in Fig.~\ref{fig:fg:transparency}, we provide a representative selection of four requirements of each category; see the supplementary materials~\cite{norval2023git} for the full set.

In all, this analysis gives us a broad set of categories reflecting user-derived requirements for bringing about greater levels of transparency regarding the operation of smart homes.
Looking at which requirements were more closely associated with the `Must', `Should', or `Could' of the MoSCoW priority system, we (anecdotally) observed some patterns whereby those categorised as `must' tended to focus on system functionality, such as specific features that the system should facilitate, while those prioritised as `should' often appeared more presentation-oriented, such as the use of terminology, data visualisations, and support tools.
This may perhaps be because there are many different ways in which information can be communicated to users (c.f. functionality), however, future research may be able to probe further into this.

\begin{figure*}[!t]
\centering
\includegraphics[width=.85\textwidth]{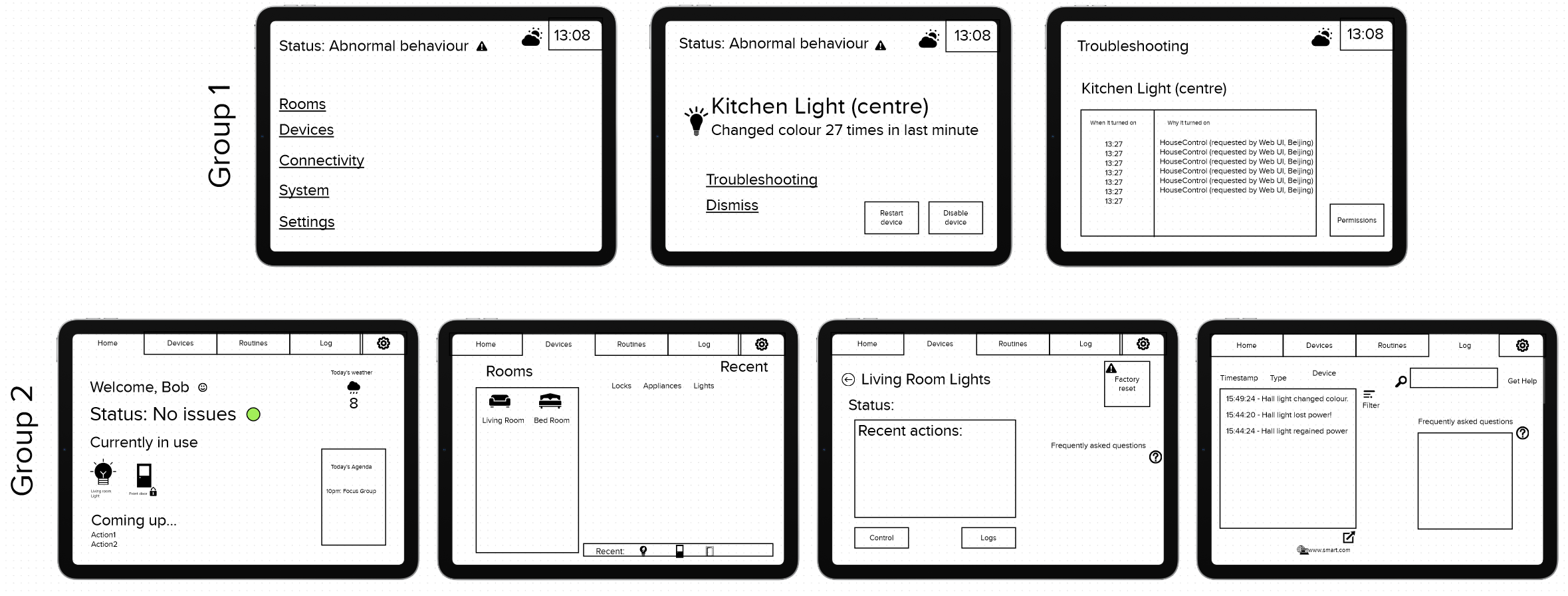}
\caption{The `Misbehaving smart bulb' scenario. Top: Group 1's interface; Bottom: Group 2's interface.\\}
\label{fig:fg:scenario1}

\vspace{.5cm}

\includegraphics[width=.85\textwidth]{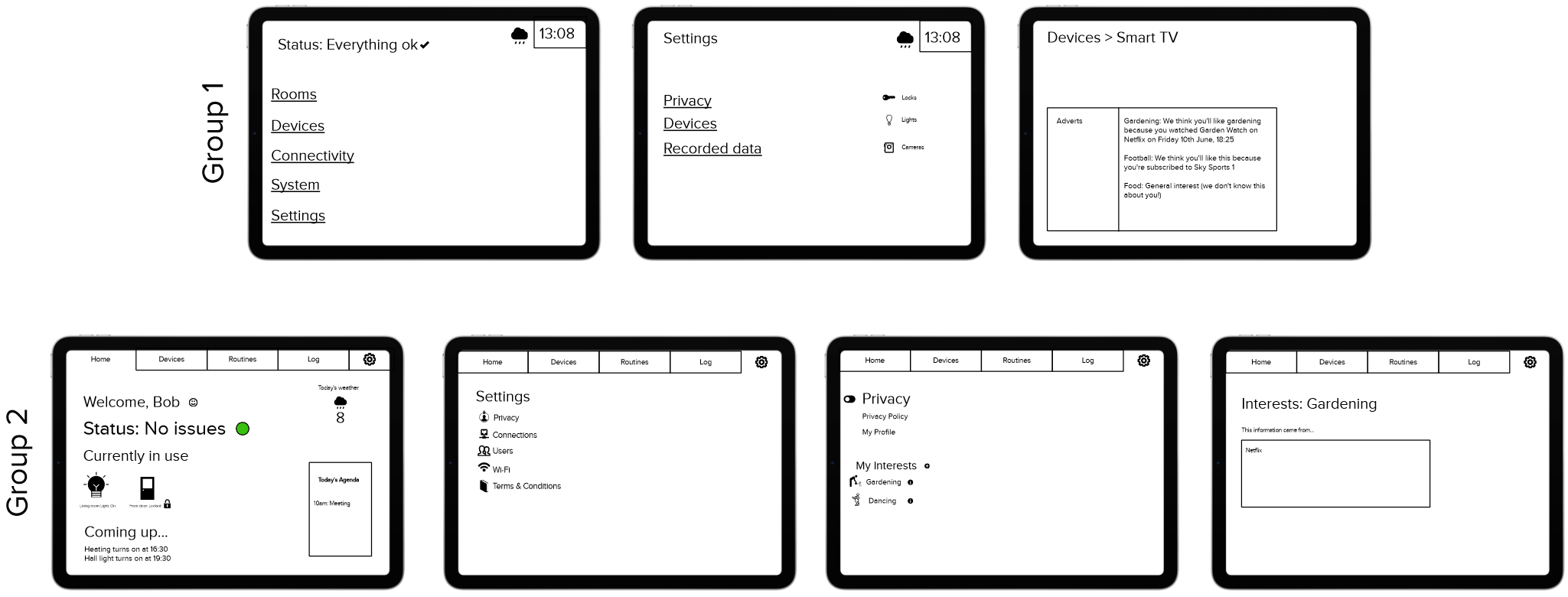}
\caption{The `Investigating adverts' scenario. Top: Group 1's interface; Bottom: Group 2's interface.\vspace{.17cm}}
\label{fig:fg:scenario2}
\end{figure*}

\begin{table*}[t]
    \centering
    \caption{The design elements that emerged from the participants' prototypes.}
    \label{tab:elements}
    \footnotesize
    \begin{tabularx}{\textwidth}{ l X }
    \hline 
    \textbf{Elements of design} & \textbf{Description} \\ 
    \hline
     \textbf{Status indication} & Both groups created a `Status' feature on the homepage, which presented information about the system's operational capabilities. This acted as a means to notify the user whenever potential issues were identified by the system, and directed them to either more information or tools to assist in understanding and rectifying the problem. In Group 1's case, the indicator entailed a direct route for navigating directly to details of the issue, whereas it appeared more of a prompt for Group 2 to dig into the interface and identify the issue at hand. \\ 

     \hline 

     \textbf{Devices list} & Both groups also had a `Devices' list, which presented a list of devices currently deployed within the smart home. This was used by Group 2 to navigate to the smart bulb during the first scenario. While not fully elaborated by Group 1 (given that they used a different means to navigate in both scenarios), they nevertheless included it as an option on the home screen. Means of navigating to the correct devices will likely be important in a transparent smart home context (particularly where several devices may be incorporated), and this approach is one potential way of enabling navigation in smart home contexts.  \\

     \hline

     \textbf{Rooms list} & Similarly, both groups had a way of navigating through the physical spaces (i.e. rooms) of the house in which their smart devices were configured and deployed. That is, both prototypes allowed the user to group and explore devices in a manner relating to the environment's physical layout. This was part of the Devices list for Group 2 but was its own menu option for Group 1.  \\

     \hline

     \textbf{Privacy settings} & Again, both groups had a Settings option on the homepage, and both settings pages prominently featured a `Privacy' sub-option. While our scenarios did not lead to either group further elaborating on what such a privacy settings page might contain, both groups listed the privacy options at the top of the list in their Settings page -- perhaps indicating the importance that the participants placed on privacy and the means for its management and control. \\

     \hline

     \textbf{Viewable system logs} & Both groups had options for viewing system logs, which played a prominent role in investigating the first scenarios. This is interesting, particularly given that much of what the participants had argued for and discussed during the sessions related to more user-friendly ways of understanding a system's operation (cf. system logs). However, this nevertheless emphasises the contextual nature of transparency mechanisms, in that both `lay' explanations, as well as more technical systems logs, may each be useful across different contexts and scenarios. Furthermore, the exact presentation of the system logs differed -- Group 1's was slightly more tabular in nature, whereas Group 2 (the students) opted for more of a user-facing explanation. Yet, much of the information presented in the interface was the same for both groups. \\

     \hline

     \textbf{Connectivity records} & Similarly, the groups both included features relating to device connections, connectivity, and any internal\slash external interactions with other devices or services. While this was stored within the `Settings' menu for Group 2, Group 1 had this menu option as a prominent option on their front page. Given the nature of our scenarios, this design element did not end up getting elaborated through either group's storyboards, but their inclusion in both interfaces indicates the potential importance of this feature. \\

     \hline
     
     \textbf{Advertising inferences list} & Both groups also outlined a page that would show advertising information (e.g. the inferences and profiles generated about the user, and how they were being used). In both cases, this page was accessible through the system's Settings page, and outlined a means to oversee and control aspects of the advertising profile that had been created within the smart home systems. \\
     
     \hline
    \end{tabularx}
\end{table*}

\subsection{Activity 2: Co-designing transparency prototypes} \label{subsec:fg:interface}

The second activity involved tasking participants with creating a prototype for a `transparency interface': a tablet/wall panel-based system, which interacted with devices within a smart home to provide greater transparency over how the devices were operating, what they were doing, etc.
This acts as a means for analysing and exploring the \textit{various types of transparency-related elements of design}  that the participants created.
That is, much like our process of identifying the {categories} of requirements, we use these prototype designs to understand more about key aspects of system design for enabling smart home transparency that our participants' interfaces contained, and how the participants perceived the design of such features in practice, so to derive the design elements. 
Again, this activity took place via MURAL, using simple shapes, text, and icons to put together a simple set of storyboards showing what features such a system could have and how it would be used to investigate and explore the operation of smart homes. 

The participants began by creating the prototype's home screen; that which would be seen when initially interacting with the tablet or wall panel.
Participants were at liberty to add components (icons, text, shapes, etc.) to the prototype as they saw fit, and the process was chaired (i.e. moderated) by the researcher. 
Once the participants had created the home screen, we then performed two sub-activities in turn, each relating to a hypothetical scenario where something in the smart home warranted attention or investigation.
The participants were first presented one of these scenarios, tasked with creating the subsequent screens that would `storyboard' how their prototype could be used to investigate the scenario.
They then repeated the process on the second scenario.
In all, these activities helped to further derive a set of transparency design elements from the participants' prototypes, while providing a context to show how their systems might work to enable meaningful transparency from their perspectives.

The two scenarios were designed to explore and reflect real-world concerns (elaborated below), which work to ground and contextualise the activity for the participants.
Both scenarios were formulated such that there could have been a number of potential underlying reasons for the concern, and a number of ways in which someone could use transparency mechanisms to investigate. 
As the participants designed new screens that would allow them to dive deeper into their prototypes, the researcher provided more information as to the actual nature of what was happening within the prototype.
For example, when participants were designing the Troubleshooting (Group 1) and Log (Group 2) pages, the researcher informed them as to what these logs would report (remote requests were originating from overseas), as a way to simulate the discovery and diagnosis process for a previously unknown issue (a staged approach, similar to that used in Study 1; \S\ref{sec:survey:questionnaire}).
In this way, we gathered information about what design elements the participants expected such a system to have, what these features and pages within the system might look like, and how they would expect to be supported in eventually finding the information that they were looking for.
In all, this design exercise produces a rich set of insights into how our participants thought transparency mechanisms might better work to inform.

\textbf{Scenario 1 -- Misbehaving smart bulb:} 
The first scenario was security-related, featuring a smart bulb that was being acted upon by a malicious actor (hijacking a smart bulb appears a common exemplar~\cite{cuthbertson2020, ronen2016, ronen2017}).
The participants were told that the bulb starts to behave strangely, changing colours seemingly unprompted and occasionally flashing rapidly, and were then tasked with co-designing the steps that they would take on their interface, to investigate this issue further.

\textbf{Scenario 2 -- Investigating adverts}
The second scenario was privacy-related, and involved suspected data leaks being used for targeted advertising (a real concern that many people have~\cite{chalhoub2020c}, and that has been observed happening in practice~\cite{avast2021, mandalari2021}).
This involved participants being told that their recently purchased smart doorbell (with a camera) had been placed on their porch overlooking their front garden, and at around the same time, they started receiving personalised adverts on their smart TV for gardening.
Given that the camera would have seen them gardening, the implication is whether their new device might be invading their privacy, by sharing or leaking information about their gardening habits with the TV's advertising provider.
As the participants created the prototype interface screens, these would reveal (steered by the researcher) that the doorbell was not involved in the gardening adverts; rather, the gardening inference was made as a result of i) their smart TV's viewing history of gardening shows, and ii) online purchases for gardening equipment.
As such, the scenario was designed to represent a case where initial suspicions led to further investigation, revealing that the system was operating in a different way from what was originally expected.
Also in contrasting with the first, the second scenario did not relate to a specific technical issue or incident; rather it involved detailing the steps that the participants would expect to take more broadly, to diagnose and understand what was occurring within their smart home through a targeted investigation.

\subsubsection{Activity 2 Findings -- Design elements for transparent smart homes} \label{sec:design_elements}

This activity resulted in two prototype systems (one produced per group), each with two `storyboards'~\cite{truong2006} (for each scenario).
These storyboards illustrated how the user would engage with the transparency interface to investigate and understand what was happening within their smart home.
See Fig.~\ref{fig:fg:scenario1} for the storyboards for the first scenario, and Fig.~\ref{fig:fg:scenario2} for the second scenario).

Note that the two groups (and, by extension, the two workshops) were run independently, in that neither group saw the prototypes, requirements, or any other outputs from the other. 
Interestingly, however, the two sets of storyboards appeared to show a number of similarities and overlaps between the two groups in term of their transparency-oriented design feature and functionality, suggesting that these overlaps may be more generally applicable.
As such, we opted to thematically categorise aspects of the prototypes' interfaces (much like we categorised responses in our earlier survey, and categorised the types of requirements in Activity 1), in order to explore this concept further.
We call these categories of interface components \textit{design elements}.

This involved thematically grouping aspects of the design, features, and functionality as they concerned the transparency mechanisms prototyped by our participants. 
For example, both groups' prototypes contained a `Devices' page or tab, which presented the user with a list of devices that were deployed within the smart home.
This resulted in a `Devices List' design element, which we then created a description for, based on how the groups had implemented it.

As Table~\ref{tab:elements} presents, there were a number of common transparency-related design elements that both groups had independently devised. 
As such, this list represents a set of corroborated, user-derived design considerations relevant for those designing and/or developing smart home systems. 
Note we discuss the implications of some of the more unique design decisions (i.e. those incorporated by only one group) in \S\ref{subsec:limitations}.

\subsection{Summary}

Through these two co-design workshops, participants' \textit{perspectives, needs, and expectations} were explored with regard to how they believed that transparency mechanisms for smart home technologies should be brought about in practice.
These workshops resulted in range of \textit{user-defined} insights toward the types of features and functionality that might help bring about greater levels of transparency within the consumer IoT. 
We discovered nine groups of requirements---five related directly to transparency-related concerns (Fig.~\ref{fig:fg:transparency}), and four concerned broader controls and mechanisms that the smart home should support (Fig.~\ref{fig:fg:wider_reqs}). 
We also derived seven design elements (Table~\ref{tab:elements}) representing the key aspects of system design (that both groups had outlined in their prototypes) for enabling smart home transparency.
These outcomes provide us with tangible ways forward for the design and implementation of meaningful transparency mechanisms within the consumer IoT -- as envisioned by our participants.

\begin{figure*}[!t]
\centering
\includegraphics[width=.85\textwidth]{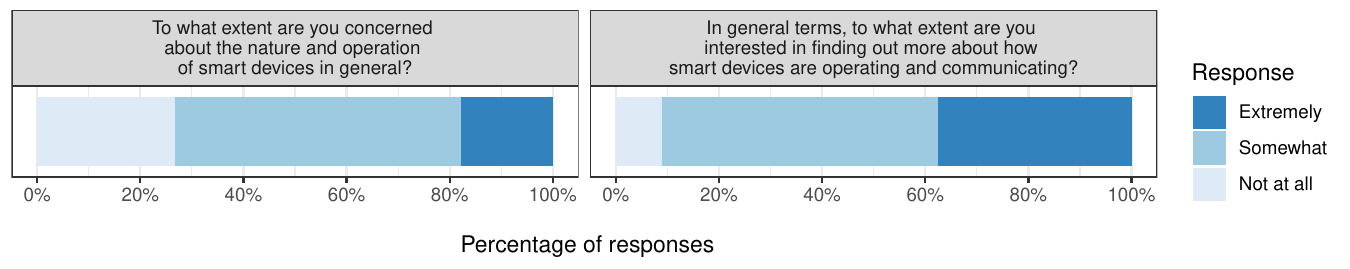}
\caption{Again, the majority of respondents from Study 3 were also concerned about the nature and operation of smart devices, and interested in finding out more about them.}
\label{fig:eval:concerned}
\end{figure*}

\section{Study 3: Validating the efficacy of our design elements and transparency requirements} \label{sec:eval}

Through the previous studies, we have derived i) a selection of categories, representing the types of user requirements that our participants raised, and ii) a collection of design elements, offering practical ways forward for developers and designers wishing to bring about improved transparency within their IoT products.
Here, we describe a further study to validate whether i) the design elements appear usable and effective in supporting meaningful transparency, and ii) the perceived relevance and general coverage of the types of transparency requirements that we have identified.

To explore whether the applicability of our categories of user requirements and design elements appeared to carry forward, we recruited 56 new participants from MTurk to take part in this questionnaire (survey).
While we used MTurk filtering criteria to ensure that none had taken part in our earlier studies (\S\ref{sec:survey:questionnaire}), our new cohort had similar characteristics to that of Study 1;
participants had similar demographics, expressed a range of technical expertise, and overwhelmingly had smart devices in their homes (as shown in Table~\ref{tab:demo3}).
Furthermore, {by chance,} when asked the extent to which they were concerned about the nature and operation of smart devices, and whether they were interested in finding out more about how they were operating and communicating (as we did for the participants in Study 1), our new cohort responded with similar characteristics (shown in Fig.~\ref{fig:eval:concerned}) to those of Study 1.
That is, the samples of both surveys appeared similar (in terms of demographics, technical expertise, in owning IoT devices, and general concerns and interests relating to smart home transparency).

\begin{table}[hbt]
    \centering
    \scriptsize
    \caption{Key demographic information of the participants from the third study.}
    \begin{tabular}{ l c }
    \hline
    ~ & \textbf{\% of respondents}\\
    \hline
     Gender & ~ \\ 
     \quad Female & 36\% \\  
     \quad Male & 64\% \\
     \quad Other & 0\% \\
     \hline
     Age & ~ \\ 
     \quad 18--29 & 38\% \\  
     \quad 30--39 & 43\% \\
     \quad 40--49 & 14\% \\
     \quad 50--59 & 4\% \\
     \quad 60+ & 2\% \\
     \hline
     Technical Expertise & ~ \\ 
     \quad No knowledge & 4\% \\  
     \quad Some knowledge & 18\% \\
     \quad Average level of knowledge & 38\% \\
     \quad Advanced knowledge & 36\% \\
     \quad Expert knowledge & 5\% \\
     \hline
     Knowledge of Smart Devices & ~ \\ 
     \quad No knowledge & 2\% \\  
     \quad Some knowledge & 11\% \\
     \quad Average level of knowledge & 48\% \\
     \quad Advanced knowledge & 32\% \\
     \quad Expert knowledge & 5\% \\
     \hline
     Have Smart Devices in the Home & ~ \\ 
     \quad None & 2\% \\  
     \quad One or more & 98\% \\
     
     \hline
    \end{tabular}
    
    \label{tab:demo3}
\end{table}

\subsection{Activity 1: Exploring the efficacy of the design elements}

\begin{figure*}[t]
\centering
\includegraphics[width=0.85\textwidth]{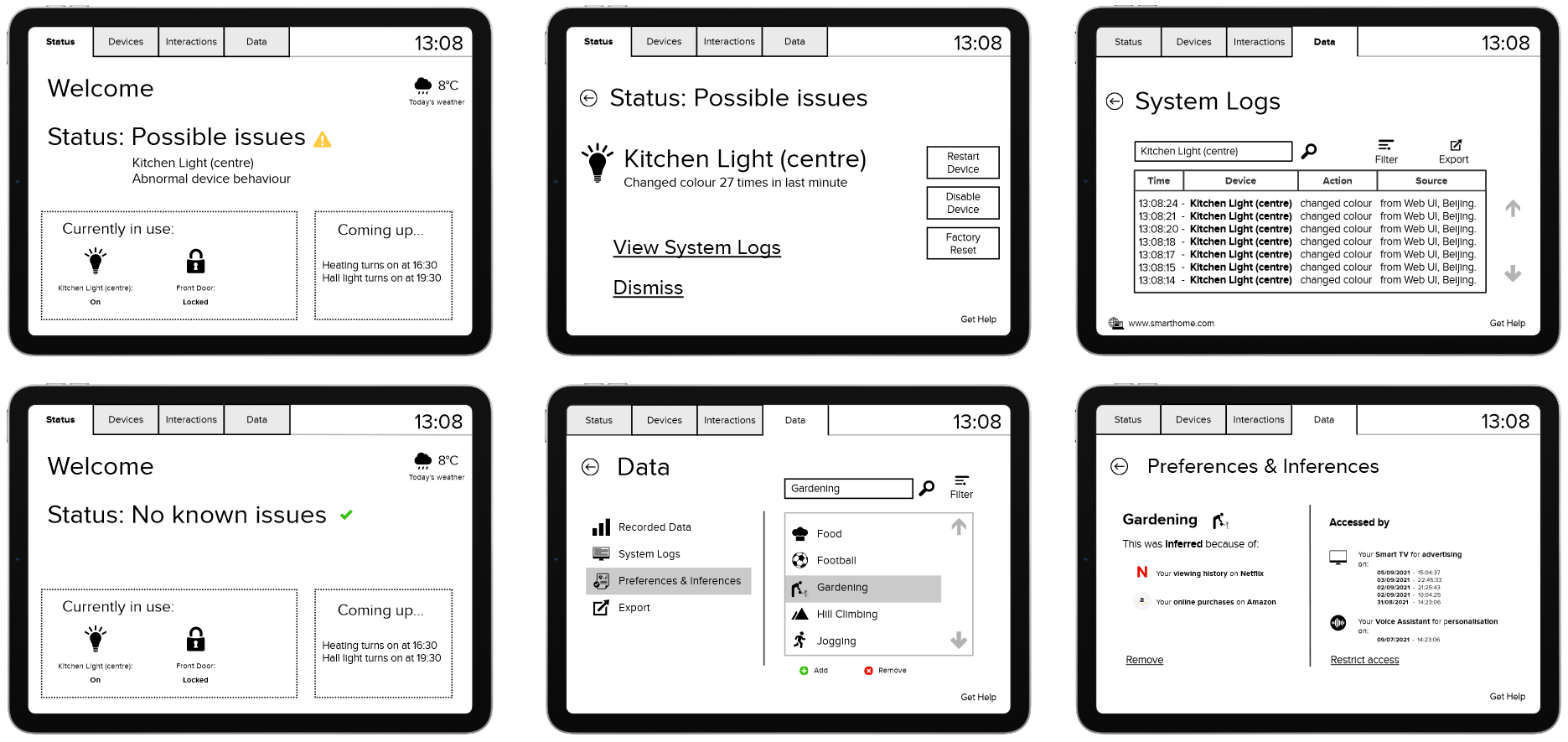}
\caption{The `merged' interface, combining the design elements from both groups' prototypes. Top: Scenario 1; Bottom: Scenario 2.}
\label{fig:fg:merged}
\end{figure*}

We begin by considering the design elements derived from the workshops of the prior study. 
Our focus is exploring whether and how these design elements assisted meaningful transparency through communicating relevant information to these new participants.
Given the similarities of the two groups' interfaces from the design study (\S\ref{subsec:fg:interface}), we opted to combine the key design elements (identified from \S\ref{sec:design_elements}) into one interface. 
In other words, we merged the key features of each interface, while maintaining a consistent and coherent design throughout (so as not to distract participants with aesthetic considerations).
The result was a new set of storyboards (i.e. a new interface; see Fig.~\ref{fig:fg:merged}) with which we could explore the efficacy of the common design elements that emerged.

After agreeing to take part, participants were first presented with a broad description of one of the scenarios (randomised order) from \S\ref{subsec:fg:interface}, and each storyboard image corresponding to that scenario was then shown to the participant, alongside textual descriptions of each `step' in the process. 
Again, this reflected the investigative process of using such an app, taking an exploratory approach to illustrate how one might use the system to investigate the scenario at hand.

After the first scenario's storyboard was presented to the participant, they were then asked (via open-text boxes) what had happened; those that started with Scenario 1 were asked why the bulb was acting erratically, whereas those that started with Scenario 2 were tasked with reporting whether the doorbell was involved in the gardening adverts.
We then asked a follow-up question relating to how confident they were in their previous answer (3-point Likert), alongside a further open text box where they could elaborate on their confidence.
By asking these questions, we probe the interface's ability to both \textit{convey the appropriate information}, and how \textit{confident the users felt} that they understood what happened -- both key aspects with regard to meaningful transparency.
This process was then repeated with the other scenario, exposing the participant to how the aforementioned design elements could assist in investigating both particular issues or concerns, while mitigating for order effect biases~\cite{perreault1975}.

After completing both scenarios, participants were then asked to complete a System Usability Scale survey (SUS)~\cite{Brooke1996} about the prototype.
The SUS comprises a set of ten Likert questions on aspects such as ease of use, consistency, and complexity, and can be used to generate a 0-100 score representing ``a general quality of the appropriateness to a purpose of any particular artefact''~\cite{Brooke1996}.
In this way, we use the SUS to gauge the extent to which the participants found the prototype to be appropriate, or usable, as a means for engaging with the transparency mechanisms and understanding what happened within the system.

Exploring i) what participants thought happened in the system, ii) how confident they were, and iii) whether the system received a `good' SUS score (indicating general aspects of system usability), provides three metrics about the extent to which the prototype supported participants in meaningfully interrogating the information at hand.

\subsubsection{Activity 1 Findings -- Exploring the potential of design elements} \label{sec:eval:act1:findings}

{To explore whether the design elements helped in enabling meaningful transparency, we look to see what our participants thought what was happening within the two scenarios.}
For the first scenario, 66\% of participants correctly listed a security breach as the potential reason for the bulb changing colour, and 91\% of participants correctly identified the source of the adverts in the second scenario (Table~\ref{tab:survey3_design_elements}).
It is worth noting that, similar to the responses of \S\ref{sec:survey:findings}, these answers were evaluated based on open-ended text, as opposed to selecting from multiple options, etc. 
That is, $\sim$two-thirds of responses specifically deduced that the requests coming from overseas was the result of a security breach.
This is interesting, showing that while a majority of participants correctly identified that a security breach was to blame, a non-negligible proportion did not (with the remaining $\sim$third typically suggesting technical faults or other issues more generic). 
This, again, shows the importance of assisting users in interpreting the information, and that different users may require varying levels of support in doing so.

\begin{table}[h]
    \centering
    \caption{Results from Study 3, Activity 1. Participants indicated (via open text answers) what they thought was the underlying reason for each scenario, and their confidence in the answer.}
    \label{tab:survey3_design_elements}
    \scriptsize
    \begin{tabular}{ l r }
    \hline
    
     \textbf{Scenario 1: Misbehaving smart bulb} & ~\vspace{.05cm} \\ 
     \quad \textbf{Correct} (n = 37) & 66.1\% \vspace{.05cm} \\

     \quad Extremely confident & 30.4\% \\ 
     \quad Somewhat confident & 58.9\% \\
     \quad Not at all confident & 10.7\% \\
     
     \hline 

    \textbf{Scenario 2: Investigating adverts} & ~\vspace{.05cm} \\ 
     \quad \textbf{Correct} (n = 51) & 91.1\% \vspace{.05cm} \\ 

     \quad Extremely confident & 41.1\% \\ 
     \quad Somewhat confident & 55.4\% \\
     \quad Not at all confident & 3.6\% \\
     
     \hline
    \end{tabular}
\end{table}

Also relevant is the degree of confidence that users felt about their answers, as shown in Table~\ref{tab:survey3_design_elements}. 
Interestingly, the two sets of proportions appear somewhat similar, despite proportion of correct responses being different; while ensuring user confidence is high is undoubtedly a key aim, it is worth recognising that those who are extremely confident in their interpretation of smart home data won't always be correct in their answers. 
In other words, while having transparency mechanisms that inspire confidence in users' understanding over how their smart homes operate is important, so too is working with various types of users---across different use cases and contexts---to ensure that these work to speak to users and their levels of expertise, and indeed, work to inform and empower, rather than mislead or oppress.

We then look to the SUS scores as another indicator as to the design elements' potential for facilitating meaningful transparency.
Low SUS scores might indicate that the participants did not find the resulting functionality particularly useful or usable. In contrast, higher SUS scores indicate a generally more `appropriate'~\cite{Brooke1996} system for the task at hand (interrogating the transparency information to determine what had happened). 
Prior work has determined that a ``poor'' SUS score is around 35.7, ``okay'' is 50.9, and ``good'' is 71.4~\cite{bangor2009}.
Our merged interface received an mean SUS score of 72.1 ($\sigma$ = 16), faring well according to these SUS benchmarks. 
Again, the SUS explores aspects such as consistency, complexity, ease of use, and perceived confidence in using the system, and offers several relevant dimensions toward enabling meaningful transparency (\S\ref{sec:meaningful_transparency}).

\subsection{Activity 2: Evaluating the coverage and relevance of transparency requirements}

We next consider the types (or `categories') of requirements that were identified in \S\ref{sec:fg:reqs}. 
To recap, the co-designed requirements that were created in the second study were categorised into nine types; five related directly to transparency-related concerns (allowing the user to `oversee', `explore', and `investigate' the transparency information, for the system to `notify' the user when necessary, and several features relating to `usability'; Fig.~\ref{fig:fg:transparency}), whereas the remaining four concerned broader controls and mechanisms that the smart home should support (allowing the users to retain `control', `safety', and `security' with regards to their smart home, with additional suggestions for user `engagement'; Fig.~\ref{fig:fg:wider_reqs}).
Given that our focus is on how we might bring about meaningful transparency mechanisms in practice, we focus on the former---those transparency-related---and explore the extent to which these appear representative and applicable to our new participants.

This activity began by presenting participants with a brief summary of each of the five transparency-related categories of requirements (see Fig.~\ref{fig:fg:transparency}). 
This contained the five categories (oversee, notify, explore, investigate, and usability), alongside a short description and examples for each.
Participants were first asked to prioritise these categories in terms of importance through allocating 100 points across each of these five categories (with more points representing higher importance).
We also asked (via open-ended text) whether there were other categories or specific requirements that the participant thought were absent, but should be included.
These questions allowed us to gain a sense of priority over these categories, and broadly to what extent they covered the transparency concerns or interests that the participants could identify.

Participants were then presented, in turn, three of the five categories (randomly selected, and presented in randomised order) alongside questions relating to the category in question.
These questions asked for participants' general thoughts regarding that category; other examples of requirements that they thought would fit into this category; and how the participant thought those requirements could best be illustrated or implemented in practice.
Again, these questions help to build up our understanding of the five categories that emerged from the co-design workshop and their general applicability across a wider sample.

\subsubsection{Activity 2 Findings -- Exploring the requirements' coverage} \label{sec:eval:reqresults}

We start by looking at the prioritisation of requirements. 
Recall that participants are tasked with allocating 100 points across the five categories (which would result in 20 points for each category, if all were seen as equally as important).
Each of these five categories appeared to receive similar scores (as seen in Table~\ref{tab:survey3_requirement_categories_prio}), and these proportions did not appear statistically different to each other.
As such, it appears as though our five categories were all seen as broadly equal in importance by our participants, with no category appearing significantly more or less important (though further research with larger sample sizes could work to investigate whether and how these differences may manifest).
In other words, each of our five categories appear to show value toward understanding the types of requirements that IoT developers may wish to aim for.

\begin{table}[!t]
    \centering
    \caption{Users were asked to allocate 100 points across each of the five categories of requirements. By default, these were distributed evenly, with each category receiving 20 points.}
    \label{tab:survey3_requirement_categories_prio}
    \scriptsize
    \begin{tabular}{ l c c }
    \hline
    
     \textbf{Category} & \textbf{$\bar{x}$} & \textbf{$\sigma$} \\ 
     \hline 
     
     \textbf{Oversee} & 20.3 & 9.4 \\ 
     \textbf{Explore} & 15.1 & 7.0 \\ 
     \textbf{Investigate} & 18.4 & 8.2 \\ 
     \textbf{Notify} & 24.1 & 9.6 \\ 
     \textbf{Usability} & 22.2 & 13.0 \\ 
     
     \hline
    \end{tabular}
\end{table}

When asked if there were any other categories of transparency requirements that the participants thought were missing or should be included, 23.7\% (n = 18) of responses included a suggestion (all of which are included within the supplementary materials).
However, when using thematic analysis~\cite{braun2006} to categorise these responses, all of the suggestions appeared to fall into our existing categories (i.e. those shown in Fig.~\ref{fig:fg:transparency} and Fig.~\ref{fig:fg:wider_reqs}). 
For example, one participant suggested ``a category dedicated to protecting my smart home from hackers'' (i.e. `Security'); whereas another asked for ``info on data is a big one and controlling who it's shared with'' (`Control').
These findings are again interesting, given that they appear to corroborate the types of requirements and concerns being raised by our participants of Study 2, and suggest the coverage of our requirement-types is representative.

\subsection{Summary}

In all, these findings appear promising; they demonstrate: i) that the design elements (realised through the prototypes) appear to have potential in supporting users in meaningfully interrogating transparency information -- demonstrated through a large proportion of participants correctly, and confidently, determining what was going on within the smart home; ii) that the merged interface was considered to have a `good'~\cite{bangor2009} amount of usability, receiving an average SUS score of 72.1 (out of 100); and iii) that the categories of requirements appear to represent the transparency-related interests and concerns of an altogether new set of participants, finding that no new categories emerged from the exercise, and demonstrating the broad coverage, relevance, and applicability of these requirement types.

These findings offer a foundation for understanding how meaningful transparency might be achieved within smart homes (and, indeed, beyond).
Next, we elaborate some of the implications of our work, and of transparency in the consumer IoT more broadly.

\section{Discussion} \label{sec:discussion}

Earlier, we described how there is a clear need for work which attempts to understand how we might bring about improved transparency mechanisms within a consumer IoT context, and the importance of working with prospective users throughout the process (\S\ref{sec:meaningful_transparency}). 
Towards this, we have undertaken a set of user studies that i) demonstrate the appetite for greater levels of transparency surrounding consumer IoT deployments (\S\ref{sec:survey}); ii) identify paths forward toward the practical development of more meaningful transparency mechanisms, through understanding more about (a) the types of requirements and (b) design elements that our participants felt that smart homes should provide (\S\ref{sec:focusgroups}); and iii) validate the coverage of these types of requirements, and the efficacy of these design elements, with an altogether new set of participants (\S\ref{sec:eval}).
Our findings provide practical insights for IoT developers and researchers alike, toward enabling more meaningful transparency mechanisms within the consumer IoT.
In realising more effective transparency mechanisms, we can help to support scrutiny, and thereby accountability -- a concern which will only grow in importance given increasing consumer demand and emerging regulatory requirements for such.
As such, we now discuss some of the broader aspects of our work.

\subsection{The importance of meaningful transparency} \label{subsec:disc:need}

Across our three studies, we have demonstrated the clear desire that users have for greater levels of transparency within the consumer IoT.
This was found rather explicitly in both the first and third study, where the vast majority of participants were somewhat or extremely concerned about the nature and operation of smart devices and environments, and were interested in finding out more about how their IoT systems were operating. 
Such findings demonstrate \textit{the importance that many of our participants placed upon transparency mechanisms}, and their role in bringing about an IoT that more closely aligns with their needs and expectations.

However, as discussed, it is well-established that simply `dumping' information on users will not necessarily be effective~\cite{acquisti2013, bovens2007, obar2020, stohl2016, suzor2019}.
Indeed, this was recognised by some participants, for example, with one warning of the risks of \textit{``information overload"}.
Furthermore, also crucial is that transparency mechanisms do not work to mislead, distract, or to otherwise provide users with artificially inflated levels of confidence~\cite{norval2022} (as alluded to in \S\ref{sec:eval:act1:findings}).
Therefore, careful consideration into the design and evaluation of transparency mechanisms will likely be of the utmost importance.

It is for this reason that we have emphasised the importance of transparency mechanisms that are \textit{meaningful} for users; such that they directly cater to the needs, requirements, experiences, and levels of expertise of a broad range of people~\cite{norval2022}. 
Towards this, our research has focused on elucidating the transparency mechanisms that \textit{the participants themselves} felt were important, and how such mechanisms could better work to support their aims and interests within a smart home context.
In doing so, we present the types of user requirements and design elements that our participants derived, which they thought would better allow them to understand their smart environments and to support them in taking action in response when necessary.

Furthermore, our findings appear to indicate a broader consensus -- not only over the importance of meaningful transparency, but also with what designing for meaningful transparency might mean within a smart home context.
This consensus was demonstrated through our validation study (Study 3), and was also apparent from our co-design workshops, where we observed considerable overlaps between the outcomes of our two (independent) groups of participants.
In this way, our findings may represent a promising starting point for illustrating the importance of meaningful transparency within smart homes, and how users might be better supported in understanding why, and how, these systems operate in the way that they do.

\subsection{Next steps and future research opportunities} \label{subsec:limitations}

Our findings represent a starting point for understanding how transparency mechanisms might better meet the needs and expectations of users.
While our categories of requirements and design elements offer tangible ways forward in this regard, our results are intended as indicative; our goal is not to argue that our findings comprise a fully representative set of requirements, design elements, or considerations that might exist in the wider population. 
Rather, our findings reflect how our participants believed that transparency mechanisms for smart homes might be achieved in practice.
In this way, our research provides a foundation for developers and researchers alike to consider, use and build upon.
Towards this, we now identify a few areas where future research might be able to assist.

\subsubsection{Research methods and contexts}

Our aim was to document and explore the types of requirements and interface components that our participants thought would assist transparency. 
Future work could explore the deployment of these outcomes `in the wild', with actual consumer IoT devices and users. 
This might entail building upon the requirements and design elements presented here, perhaps exploring how best our outcomes can be translated across the various contexts and scenarios that might arise in any given IoT deployment.

Importantly, however, is that issues of transparency are contextual, and what is needed will often depend on circumstances.  
Recall that our studies were scenario-led (\S\ref{sec:survey:questionnaire}, \S\ref{subsec:fg:interface}); while these scenarios reflected grounded concerns and interests that real people have with the consumer IoT (\S\ref{subsec:methodology_grounding}), future research may consider exploring different scenarios and contexts. 
Similarly, while we mainly focused on the context of a `control panel' for consumer smart homes (\S\ref{subsec:fg:interface}), and thus, outlined elements suitable for such a modality, there are many ways that transparency-related information can be presented and communicated. 
In all, while these decisions provided some necessary scoping and grounding for our research studies, there are many additional research opportunities, such as those focusing on different scenarios and use cases, different types of transparency mechanisms, system modalities, interaction techniques, and other types of connected environments.

\subsubsection{Participant samples}

There is also scope for considering our research within different cohorts and samples of participants.
Recall that we used Mechanical Turk for recruitment to our surveys; while we placed few restrictions on who could take part (\S\ref{subsec:methodology_recruitment}), our sample did appear somewhat limited in representation (e.g. fairly `techy', with the vast majority having at least one smart device in their home).
Similarly, for our co-design workshops, we used a group of computer science undergraduates as a point of comparison.
While the use of such a `convenience sample' allowed us to recruit participants with some knowledge of system requirements and interface prototyping (as a means of ensuring that their responses were grounded within some degree of systems design), important is that we do not intend it to be reflective of the population at large.

While we did verify our findings with an altogether new set of participants (\S\ref{sec:eval}), follow-up research could explore the extent to which they generalise to new audiences, and how different samples may express different (or similar) characteristics. 
For example, note that for Study 3, we only considered the design elements that showed consensus across both groups (\S\ref{subsec:fg:interface}) -- omitting, for example, elements such as a `Settings' menu and a `Routines' tab for creating simple automated script (see Table \ref{tab:elements}). 
This indicates that there is potential for future work that explores other potential design elements, across a range of cohorts.

\subsubsection{Features and functionality}
Our work was user-focused, where participatory methods were used to have the participants themselves develop, determine, and create various transparency-related requirements and interventions. 
Naturally, there are opportunities for future work to probe on specific features and functionality related to transparency and other related issues. 
For example, it may be useful to explore  particular methods for privacy preservation, including techniques to perturb or obfuscate data, and methods to restrict unintended data access~\cite{loukil2017}.
Similarly, exploring specific means for meaningfully describing the purposes for which smart home data is collected, used and transferred is an area for further exploration. 
Indeed, this could entail probing or adapting various approaches from literature, such as providing descriptions of apps and devices~\cite{crabtree2018}, creating data visualisations for sensor feeds~\cite{castelli2017}, or `nutrition labels'~\cite{railean2018, railean2021}.
While out of scope for this particular paper, exploring how participants might look to influence the design of such approaches might represent promising areas for future research.

More broadly, given the importance of transparency mechanisms (and the oversight, scrutiny, and accountability that they can enable), there is a clear need for research which furthers our understanding of the risks and implications associated with transparency mechanisms.
This includes work on ensuring that transparency mechanisms are effective in communicating specific risks and concerns, and importantly that they operate to empower and inform, not mislead or distract. Further, work relating to concerns over malicious design practices (i.e. `dark patterns')~\cite{gray2018}, which we see being discussed across a range of technical contexts, also warrants consideration here~\cite{kowalczyk2023}.

\subsubsection{Applicability beyond the consumer IoT}

Lastly, though we have focused on the consumer IoT, our research has broader relevance;
transparency and accountability---as they relate to technologies more broadly, beyond that of the IoT---are topics of growing importance~\cite{pasquale2015}. 
This is because transparency will often be a precursor to broader accountability aims, and users seeking a greater understanding of technologies will often be doing so in response to particular issues or concerns. 
That is to say, there is also much scope for similar research that focuses on transparency mechanisms for different types of technologies (beyond  the consumer IoT).
Indeed, there are various other technologies facing calls for increased accountability---including AI and algorithmic systems~\cite{cobbe2021, singh2019}, augmented, mixed and virtual reality~\cite{cloete2020, cloete2022, norval2023}, and cloud services~\cite{cobbe2021, javadi2021, millard2021}, to name a few---and research which works with users to derive meaningful transparency mechanisms may have much to offer.
Such research might provide new contributions (some of which will transcend across the specific technologies in question), and can therefore help to further our understanding of meaningful transparency mechanisms in general, and how they can work to facilitate greater levels of oversight and understanding.

In all, our findings are but one piece of a much larger puzzle; through working with our participants to design novel transparency mechanisms for smart homes, our work lays the foundations for, and aims to bring more attention to, this nascent research area. 
There is real potential for work---across various different user groups, scenarios, contexts, and, indeed, technologies---to build upon that which we have found.
Furthermore, understanding the similarities and differences across these different samples, contexts, and technologies might help us to build a more comprehensive understanding of how we might realise more meaningful transparency mechanisms.
As such, the above represent but a few of the areas where new research has much to offer, amid the growing demands for more transparency surrounding the technologies that are becoming ever more commonplace within our lives.

\subsection{Drivers for change: Encouraging better practice} \label{sec:disc:reg}

Questions around the motivations of tech organisations, and drivers for change, are worth considering.
While one might suggest that it is  not in the tech organisations' interests to facilitate greater scrutiny into their actions, doing so can offer some advantages that organisations may wish to consider going forward. 
This is because there are growing pressures for increased transparency regarding technologies in general, and organisations themselves can also stand to benefit by being proactive in delivering transparency mechanisms that are more in line with consumer demands.
This may be, for example, for reasons of reputation (showing that they take their responsibilities as tech developers/manufacturers/services seriously), or perhaps for reasons of competitive advantage (as we have already seen raised within the context of data protection~\cite{norval2021b}).

Nevertheless, these pressures for increased transparency within the tech sector go beyond consumer demand alone; transparency demands are also increasingly arising from law (e.g. the EU's GDPR~\cite{gdpr2016}), standards bodies~\cite{winfield2021}, and civil society~\cite{eff2022, org2022}. 
Indeed, when asking workshop participants whether they thought the co-designed system would be useful, it was questioned whether governments could do more to ``require that [organisations] are open'' -- an argument that appears to be growing in prominence~\cite{chalhoub2020a, chen2021, haney2021, urquhart2020}.

In fact, issues relating to transparency and accountability are already prompting regulatory attention, and we do see transparency requirements playing a key role in a number of emerging regulatory regimes, relevant to the IoT and beyond.
These include various regulations that are emerging from the European Union, including the GDPR, the AI Act (AI is commonly used in IoT contexts to process data from sensors, enable automation through trigger actuators, and so on~\cite{kok2023}), the Internet of Things Policy~\cite{eu2022} and proposed Cyber Resilience Act~\cite{eu2022cyberresil}.
Other examples include the UK's proposed Consumer IoT Regulation~\cite{gov2021securebydesign} and Code of Practice for Security of IoT~\cite{gov2018}. 
More broadly, governments and legislators appear to be playing an increasingly prominent role in helping to encourage and define better practices (e.g. for transparency and security), and this trend of regulatory intervention looks set to continue.

Again, however, there appears a growing recognition that regulatory obligations for transparency transcend the provision of information or data alone~\cite{norval2022}.
Within the context of the GDPR, for example, representatives from EU data protection authorities have specifically recognised that ``the quality, accessibility and comprehensibility of the information is as important as the actual content of the transparency information''~\cite{wp2018}.
In other words, it is not enough that such information is `dumped' on users, and important is that they can effectively understand, engage with, and act upon the information provided through transparency mechanisms.
It is therefore apparent that work such as ours---focusing on transparency mechanisms that are \textit{meaningful and usable} for their intended users---could have a significant role to play in helping to shape better practices with regard to how such transparency mechanisms come to be expected.

\section{Concluding remarks}

As the IoT continues to proliferate, it is important to ensure that the technology operates in a manner that meets the needs and expectations of its users.
Here, transparency plays an important role -- by providing information for users about the operation of these systems, transparency mechanisms work to support consumer oversight, understanding, accountability, autonomy, and control. 
However, simply providing information can often be of limited benefit; rather, there is a need to ensure that transparency mechanisms are more effective in catering to the needs and expectations of their users.

Through a range of user studies, we provide several tangible ways forward on this under-considered topic, bringing users into the design process, and documenting how \textit{they} perceive that transparency mechanisms can best support their aims.
We uncover the types of information that prospective IoT users want to know, how they expect that to be communicated, and what follow-on actions such information might enable. 
We also present sets of participant-derived requirements and design elements for bringing about a more transparent consumer IoT, finding that these appeared to support meaningful transparency aims with an altogether new group of participants. 
That is, by giving some insight into how transparency mechanisms can better serve the needs of users within the consumer IoT, our broader aim is to help bring about more transparent and accountable technologies, and call for more attention to be brought into this important area of research.

\section*{Acknowledgment}
We acknowledge the financial support of UK Research \& Innovation (grants EP/P024394/1, EP/R033501/1), The Alan Turing Institute, and Microsoft, through the Microsoft Cloud Computing Research Centre.

\bibliographystyle{IEEEtran}
\bibliography{refs}

\end{document}